\documentclass[11pt,letter]{article}

\usepackage{amsmath,amssymb,amsfonts} 	
\usepackage{multirow}
\usepackage[pdftex]{graphicx}
\usepackage{tabularx}
\usepackage[numbers,sort&compress]{natbib}
\usepackage{hhline}
\usepackage{subfigure}
\usepackage{color}
\usepackage{url}
\usepackage[symbol]{footmisc}
\usepackage{ulem}

\DeclareFontFamily{OMX}{MnSymbolE}{}
\DeclareFontShape{OMX}{MnSymbolE}{m}{n}{
    <-6>  MnSymbolE5
   <6-7>  MnSymbolE6
   <7-8>  MnSymbolE7
   <8-9>  MnSymbolE8
   <9-10> MnSymbolE9
  <10-12> MnSymbolE10
  <12->   MnSymbolE12}{}
\DeclareSymbolFont{mnlargesymbols}{OMX}{MnSymbolE}{m}{n}
\SetSymbolFont{mnlargesymbols}{bold}{OMX}{MnSymbolE}{b}{n}
\DeclareMathDelimiter{\llangle}{\mathopen}{mnlargesymbols}{'164}{mnlargesymbols}{'164}
\DeclareMathDelimiter{\rrangle}{\mathclose}{mnlargesymbols}{'171}{mnlargesymbols}{'171}

\begin{document}

\begin{center}
{\large \bf Accurate ground state- and quasiparticle energies: beyond the RPA and GW methods with adiabatic exchange-correlation kernels}
\end{center}

{Thomas Olsen$^{1,}\footnote[1]{tolsen@fysik.dtu.dk}$, Christopher E. Patrick$^{2}$, Jefferson E. Bates$^3$, Adrienn Ruzsinszky$^4$, and Kristian S. Thygesen$^{1,5}$}

{\it\small $^1$Computational Atomic-Scale Materials Design (CAMD), Department of Physics, Technical University of Denmark}\\
{\it\small $^2$Department of Physics, University of Warwick, Coventry CV4 7L, United Kingdom}\\
{\it\small $^3$A. R. Smith Department of Chemistry and Fermentation Sciences, Appalachian State University, Boone, North Carolina 28607, United States}\\
{\it\small $^4$Department of Physics, Temple University, Philadelphia, Pennsylvania 19122, United States}\\
{\it\small $^5$Center for Nanostructured Graphene (CNG), Department of Physics, Technical University of Denmark}



\begin{abstract}
We review the theory and application of adiabatic exchange-correlation (xc-) kernels for ab initio calculations of ground state energies and quasiparticle excitations within the frameworks of the adiabatic connection fluctuation dissipation theorem and Hedin's equations, respectively. Various different xc-kernels, which are all rooted in the homogeneous electron gas, are introduced but hereafter we focus on the specific class of renormalized adiabatic kernels, in particular the rALDA and rAPBE. The kernels drastically improve the description of short-range correlations as compared to the random phase approximation (RPA), resulting in significantly better correlation energies. This effect greatly reduces the reliance on error cancellations, which is essential in RPA, and systematically improves covalent bond energies while preserving the good performance of the RPA for dispersive interactions. For quasiparticle energies, the xc-kernels account for vertex corrections that are missing in the GW self-energy. In this context, we show that the short-range correlations mainly correct the absolute band positions while the band gap is less affected in agreement with the known good performance of GW for the latter. The renormalized xc-kernels offer a rigorous extension of the RPA and GW methods with clear improvements in terms of accuracy at little extra computational cost.  
\end{abstract}



\parindent 20pt

\section{Introduction}
For decades density functional theory (DFT) has been the workhorse of first-principles materials science. Immense efforts have gone into the development of improved exchange-correlation (xc)-functionals and today hundreds of different types exists, including the generalized gradient approximations (GGA), meta GGAs, (screened) hybrid functional, Hubbard corrected functionals (LDA/GGA+U), and the non-local van der Waals density functionals. Typically, these contain several parameters that have been optimized for a particular type of problem or class of materials. Moreover they rely on fortuitous and poorly understood error cancellations. This limits the universality and predictive power of commonly applied xc-functionals and the accuracy is often highly system dependent. 

At the highest rung of the current hierarchy of xc-functionals, lie those based directly on the adiabatic-connection fluctuation-dissipation theorem (ACFDT). The ACFDT provides an exact expression for the electronic correlation energy in terms of the interacting density response function\cite{langreth1975exchange, gunnarsson1976exchange}. An attractive feature of the ACFDT is that it provides the pure correlation energy, which should then be combined with the exact exchange energy. This clear division removes the reliance on error cancellation between the exchange and correlation terms, which is significant (and uncontrolled) in the lower rung xc-functionals. A further advantage of the ACFDT is that, even in its simplest form, it captures dispersive interactions very accurately through the non-locality of the response function.  

The simplest approximation to the response function beyond the non-interacting one is the random phase approximation (RPA). The RPA generally provides an excellent account of long-range screening and it cures the pathological divergence of second-order perturbation theory for the homogeneous electron gas. However, an important shortcoming of the RPA response function is that the local ($r$ close to $r'$) correlation hole derived from it, via the ACFDT, is much too deep leading to a drastic overestimation of the absolute correlation energy by several tenths of an eV per electron. This key observation is responsible for most of the failures of the RPA and GW schemes to be discussed in this review. It occurs because the RPA response function only accounts for the Hartree component of the induced potentials. The neglected xc-component of the induced potential is short range in nature and therefore mainly influences the local shape of the correlation hole. 

Early work\cite{furche2001molecular,fuchs2002accurate,aryasetiawan2002total} applied the ACFDT-RPA to compute the dissociation energies of small molecules finding a systematic tendency of the RPA to underbind and generally lower accuracy than the generalized gradient approximations (GGA). It was also demonstrated that RPA accounts well for strong static correlation and correctly describes the dissociation curve of the N$_2$ molecule. Around ten years later, the RPA was applied to calculate cohesive energies of solids\cite{Harl2010a} again finding that RPA performs significantly worse than GGA with a systematic tendency to underbind. In contrast, RPA was found to produce excellent results for structural parameters of solids\cite{miyake2002total,marini2006first} as well as bond energies in van der Waals systems like graphite\cite{lebegue} and noble gas solids\cite{Harl2008}, which are poorly described by semi-local approximations. In addition, for the case of graphene adsorbed on metal surfaces, where dispersive and covalent interactions are equally important, the RPA seems to be the only non-empirical method capable of providing correct potential energy curves\cite{Olsen2011, mittendorfer}. While the RPA method has many attractive features, it is clear that its poor description of short-range correlations, which results in overestimation of absolute correlation energies, and underestimation of covalent bond strengths, disqualifies it as the highly desired fully \textit{ab initio} and universally accurate total energy method. One approach to improving the RPA total energy is based on the idea of correcting the RPA self-correlation energy by including higher order exchange terms in many- body perturbation theory (MBPT) and is referred to as second-order screened exchange (SOSEX). The SOSEX correlation energy vanishes for one-electron systems and improves the accuracy of covalent bonds slightly compared to RPA. However, it deteriorates the good description of static correlation and barrier heights in the RPA\cite{ren2012random, Gruneis2009}. In addition, SOSEX scales as $N^5$ with system size and therefore comprises a significant computational challenge compared to RPA, which scales as $N^4$.

In this review we focus on a different strategy, which is based on the use of time-dependent density functional theory (TDDFT) to construct better response functions. The crucial ingredient in this theory is the xc-kernel, $f_{xc}(\mathbf r, \mathbf r',\omega)$, which is the functional derivative of the time-dependent xc-potential with respect to the density. In Ref. \cite{Lein2000} it was argued, based on ACFDT calculations for the homogeneous electron gas HEG, that the frequency dependence of the xc-kernel is of minor importance for the correlation energy, while the spatial non-locality is crucial. Moreover, it has been shown that any local approximation to the xc-kernel produces a correlation hole which diverges at the origin\cite{furche2005fluctuation}. As a consequence, the use of local xc-kernels typically result in correlation energies that are worse than those obtained with the RPA. Several non-local approximations to the xc-kernel of the HEG have been proposed. On a scale given by the error of RPA, they seem to perform very similarly for the ground state correlation energy, and therefore the present review will focus on one specific form, namely the renormalized adiabatic kernels of Refs. \cite{Olsen2012, Olsen2013a,Olsen2014a}, rAX, where X refers to a ground state xc-functional. 

We show that the use of non-local kernels largely fixes the erroneous RPA correlation hole and provides a much better description of 
short range correlations - at least for weakly correlated materials. This implies that absolute correlation energies are much better and thus the reliance on error cancellation when forming energy differences is lifted. Specifically, covalent bond energies are greatly improved while the good 
performance of RPA for dispersive interactions is preserved. The performance of the rAX kernels is further discussed for structural parameters, atomization energies of molecules, cohesive energies of solids, formation energies of metal-oxides, surface- and adsorption energies, molecular dissociation curves, static correlation, and structural phase transitions. 

Beyond total energy calculations, the renormalized kernels have also been used to incorporate vertex corrections into self-energy based quasiparticle (QP) band structure calculations\cite{schmidt2017simple}. The formal basis of such calculations is constituted by Hedin's equations, a coupled set of equations for the key quantities of a perturbative treatment of the single-particle Green's function, $G$, in terms of the screened Coulomb interaction, $W$. Within the widely used GW approximation, the vertex corrections are completely ignored. Despite this omission, the GW approximation typically yields good results for the QP band gap\cite{ hybertsen_first-principles_1985, rinke2005combining, van2006quasiparticle, shishkin2006implementation, scherpelz2016implementation, huser2013quasiparticle}. Vertex corrections evaluated form the SOSEX diagram have been found to yield some improvement of band gaps and, in particular, ionization potentials of solids\cite{ gruneis2014ionization}. This is clearly 
very satisfactory from a theoretical point of view. The drawback of this approach, however, is the high complexity of the formalism, the concomitant loss of physical transparency, and the significant computational overhead as compared to the conventional GW method. Just like the two-point xc-kernels from TDDFT provide a computationally tractable strategy to improve total energies, they can also be used to approximate vertex corrections in the electron self-energy\cite{delsole}. As for the ground state energy, the local xc-kernels perform rather badly\cite{schmidt2017simple}. Instead the renormalized kernels yield a major improvement over the GW method when it comes to ionization potentials and electron affinities, i.e. absolute band energies, as a result of the superior description of short-range correlations. This can be understood as a direct consequence of the systematic underestimation of the absolute correlation energy by the RPA. In contrast, the QP \textit{gap} is only slightly affected by the vertex because it is mainly governed by long range correlations. This in fact explains the success of the GW approximation in describing QP band gaps.

An important common feature of the ACFDT and Hedin's equations is that they are typically implemented non-selfconsistently starting from some mean field Hamiltonian. This means that the results of such calculations acquire a starting point-dependence. While the LDA and GGAs are the most widely used starting points, other xc-functionals, such as exact exchange hybrids\cite{bates2016reference,bruneval2012benchmarking} and GGA+U\cite{patrick2016hubbard,miyake2006quasiparticle,kioupakis2008g}, have also been employed. In general, it has been found, for both GW and RPA, that the results are quite insensitive to the starting point; in particular, they are much less sensitive to the initial mean field than the mean field itself, e.g. the band gap obtained with GW@GGA+U and the lattice constants and energy differences determined from RPA@GGA+U, vary much less with U as compared to the GGA+U result itself\cite{patrick2016hubbard,miyake2006quasiparticle,Olsen2017}. This is clearly a desirable effect, but does not remove the fundamentally disturbing starting point dependence. The problem arises because the natural starting point for GW and RPA would be the Hartree mean field solution, which is notoriously bad. The situation is somewhat improved by the rAX kernels because their consistent starting point would be a DFT Hamiltonian with the X-functional (or more precisely a weighted density approximation to the X-functional, see supplementary of Ref. \cite{Olsen2014a}). We note in passing that ideally the calculations should be performed self-consistently. However, this is rarely done in practice and involves other problems such as overestimated band gaps and smeared out spectral features in GW and technical difficulties associated with the self-consistent determination of the RPA optimized effective potential.  

Here we focus on the theory, implementation, and implications of physics \textit{beyond} the RPA and GW methods as described by static non-local xc-kernels from TDDFT. Consequently, we will not dwell on the RPA and GW methods themselves but refer the interested reader to one of the existing reviews on these topics\cite{ EBF12, ren2012random, aulbur2000quasiparticle, aryasetiawan1998gw, onida2002electronic}. The paper is organized as follows. In Sec.~\ref{sec:theory} we present the basic theory of ground state- and quasiparticle energy calculations based on the adiabatic connection fluctuation dissipation theorem and Hedin's equations, respectively. We introduce several non-local xc-kernels for the homogeneous electron gas (HEG) and describe a renormalization procedure for constructing non-local xc-kernels from (semi-)local xc-functionals. In Sec.~\ref{sec:implementation} we describe the numerical implementation of non-local xc-kernels including different strategies for generalizing HEG kernels to inhomogeneous densities and some aspects of $k$-point and basis set convergence. In Sec.~\ref{sec:results} we present a series of results serving to illustrate the effect and importance of the xc-kernels for both total energies and QP band structures. Specifically, we assess the performance of the rALDA and rAPBE xc-kernels for structural parameters of solids, atomization energies of covalently bonded solids and molecules, oxide formation energies, van der Waals bonding, dissociation of statically correlated atomic dimers, surface- and chemisorption energies, structural phase transitions, and QP energies of bulk and two-dimensional semiconductors. Finally, our conclusions and outlook is provided in Sec.~\ref{sec:conclusions}.

\section{Theory}\label{sec:theory}
\subsection{The adiabatic connection fluctuation-dissipation theorem}
In the Kohn-Sham (KS) scheme, a non-interacting Hamiltonian is constructed such that it has a ground state Slater determinant $|\varphi_0\rangle$, which yields the same ground state density as the true ground state wavefunction $|\psi_0\rangle$. The adiabatic connection denotes a generalization of this scheme where the Coulomb interaction $v_c$ is rescaled by $\lambda$, such that the ground state wavefunction $|\psi^\lambda_0\rangle$ reproduces the electronic density of $|\psi_0\rangle$. The procedure can be accomplished by modifying the external potential $v_{KS}^\lambda(\mathbf{r})$ and we have that $|\psi^{\lambda=1}_0\rangle=|\psi_0\rangle$ and $|\psi^{\lambda=0}_0\rangle=|\varphi_0\rangle$.

The adiabatic connection allows one to obtain a highly useful expression for the correlation energy. To begin with the Hartree-exchange-correlation energy can be written as\cite{Fiolhais2003}
\begin{align}
    E_{Hxc}&=E_{tot}-T_{KS}-v_{ext}\notag\\
           &=\langle\psi_0|T+v_c|\psi_0\rangle-\langle\varphi_0|T|\varphi_0\rangle\notag\\
           &=\langle\psi_0^\lambda|T+\lambda v_c|\psi^\lambda_0\rangle\Big|_0^1\notag\\
           &=\int_0^1d\lambda\frac{d}{d\lambda}\langle\psi_0^\lambda|T+\lambda v_c|\psi_0^\lambda\rangle\notag\\
           &=\int_0^1d\lambda\langle\psi_0^\lambda|v_c|\psi_0^\lambda\rangle,
\end{align}
where $E_{tot}$ is the total electronic ground state energy, $T_{KS}$ is Kohn-Sham kinetic energy, and $v_{ext}$ is the expectation value of the external potential. In the last quality we used the Hellmann-Feynman theorem and the fact that $\psi_\lambda$ is defined as the state that minimizes the expectation value of $T+\lambda v_c$. Inserting the second quantized form of the Coulomb interaction the expression becomes
\begin{align}
    E_{Hxc}&=\frac{1}{2}\int_0^1d\lambda\int \frac{d\mathbf{r}d\mathbf{r}'}{|\mathbf{r}-\mathbf{r}'|}\langle\Psi^\dag(\mathbf{r})\Psi^\dag(\mathbf{r}')\Psi(\mathbf{r}')\Psi(\mathbf{r})\rangle_\lambda\notag\\
    &=\frac{1}{2}\int_0^1d\lambda\int \frac{d\mathbf{r}d\mathbf{r}'}{|\mathbf{r}-\mathbf{r}'|}\Big[\langle\Psi^\dag(\mathbf{r})\Psi(\mathbf{r})\Psi^\dag(\mathbf{r}')\Psi(\mathbf{r}')\rangle_\lambda-\delta(\mathbf{r}-\mathbf{r}')\langle\hat n(\mathbf{r})\rangle_\lambda\Big].
\end{align}
where $\hat n(\mathbf{r})=\Psi^\dag(\mathbf{r})\Psi(\mathbf{r})$ and we have introduced the notation $\langle\ldots\rangle_\lambda=\langle\psi^\lambda_0|\ldots|\psi^\lambda_0\rangle$. Since the last term is independent of $\lambda$, we can get rid of it by subtracting the Hartree-exchange energy $E_{Hx}$, which is given by a similar expression with $|\psi^\lambda_0\rangle$ replaced by $|\varphi_0\rangle$. We then have
\begin{align}\label{eq:E_c_n}
    E_{c}&=\frac{1}{2}\int_0^1d\lambda\int \frac{d\mathbf{r}d\mathbf{r}'}{|\mathbf{r}-\mathbf{r}'|}\Big[\langle\hat n(\mathbf{r})\hat n(\mathbf{r}')\rangle_\lambda-\langle\hat n(\mathbf{r})\hat n(\mathbf{r}')\rangle_0\Big]
\end{align}

The density-density correlation function is closely related to the density-density response function. The retarded response at vanishing temperature is defined by
\begin{align}
\chi_\lambda(\mathbf{r},\mathbf{r}';t,t')=-i\theta(t-t')\langle[\hat n(\mathbf{r},t),\hat n(\mathbf{r}',t')]\rangle_\lambda,
\end{align}
where the expectation value is with respect to the ground state. In the frequency domain it becomes
\begin{align}
\chi_\lambda(\mathbf{r},\mathbf{r}';\omega)=\sum_{m\neq0}\Big[\frac{n_{0m}^\lambda(\mathbf{r})n_{m0}^\lambda(\mathbf{r}')}{\omega-E_{m0}+i\eta}-\frac{n_{0m}^\lambda(\mathbf{r}')n_{m0}^\lambda(\mathbf{r})}{\omega+E_{m0}+i\eta}\Big],
\end{align}
where $n_{0m}(\mathbf{r})=\langle\psi_0^\lambda|\hat n(\mathbf{r})|\psi_m^\lambda\rangle$, $E_{m0}^\lambda=E_m^\lambda-E_0^\lambda$ are the eigenvalue differences, and $\eta$ is a positive infinitesimal. It is then clear that
\begin{align}
\frac{-1}{\pi}\int_0^\infty d\omega \mathrm{Im}\chi_\lambda(\mathbf{r},\mathbf{r}';\omega)&=\sum_{m\neq0}n_{0m}^\lambda(\mathbf{r})n_{m0}^\lambda(\mathbf{r}')\notag\\
&=\langle\hat n(\mathbf{r})\hat n(\mathbf{r}')\rangle_\lambda-n(\mathbf{r})n(\mathbf{r}')\notag\\
&=\langle\delta\hat n(\mathbf{r})\delta\hat n(\mathbf{r}')\rangle_\lambda,
\end{align}
with $\delta\hat n(\mathbf{r})\equiv\hat n(\mathbf{r})-n(\mathbf{r})$. The equality is an example of a fluctuation-dissipation theorem, since it relates the imaginary (dissipative) part of the density response to the correlation between density fluctuations. 

The retarded response function only has poles in the negative imaginary half-plane and its frequency integral on a closed loop in the upper right quarter of the complex plane vanishes since $\chi\sim 1/|\omega|^2$ for $|\omega|\rightarrow\infty$. We can thus switch the integration path to the positive imaginary axis where the frequency dependence is smooth. Noting that $\chi^*(\mathbf{r},\mathbf{r}';i\omega)=\chi(\mathbf{r}',\mathbf{r};i\omega)$ we obtain
\begin{align}\label{eq:E_c_x}
E_{c}&=-\int_0^1d\lambda\int_0^\infty \frac{d\omega}{2\pi}\llangle v_c\chi_\lambda(i\omega)-v_c\chi_0(i\omega)\rrangle.
\end{align}
where $\llangle \ldots \rrangle$ indicates the trace of the two-point functions involved in the adiabatic-connection integrand.

The problem of calculating the correlation energy has thus been rephrased into finding a good approximation for the density-density response function. The simplest non-trivial approximation is the Random Phase Approximation (RPA), which can be obtained from many-body perturbation theory by assuming a non-interacting irreducible response function. Alternatively, the full (reducible) response function can be obtained from Time-Dependent Density Functional Theory (TDDFT), where it can be shown to satisfy the Dyson equation 
\begin{align}\label{eq:dyson}
    \chi^\lambda(\omega)=\chi_{KS}(\omega)+\chi_{KS}(\omega)\Big[\lambda v_c + f_{xc}^\lambda(\omega)\Big]\chi^\lambda(\omega),
\end{align}
where all quantities are functions of $\mathbf r$ and $\mathbf r'$ and integration of repeated variables is implied. $f_{xc}(\omega)$ is the temporal Fourier transform of the exchange-correlation kernel
\begin{align}\label{eq:f_xc}
    f_{xc}(\mathbf{r},\mathbf{r}',t-t')=\frac{\delta v_{xc}(\mathbf{r},t)}{\delta n(\mathbf{r}',t')}
\end{align}
and any approximation to $f_{xc}$ thus implies an approximation for the ground state correlation energy in the framework of the adiabatic-connection combined with the fluctuation dissipation theorem. In the context of TDDFT, the RPA is simply obtained by neglecting the xc-kernel when solving Eq. \eqref{eq:dyson}.

In order to calculate correlation energies from Eqs. \eqref{eq:E_c_x} and \eqref{eq:dyson} it is necessary to generalize the kernel to an arbitrary coupling strength $\lambda$. In Ref. \cite{Lein2000} it was shown that $f_{xc}^\lambda$ can be obtained from $f_{xc}$ by the rescaling
\begin{equation}
f^\lambda_{xc}(n,q,\omega) = \lambda^{-1} f_{xc}(n/\lambda^3,q/\lambda,\omega/\lambda^2).
\end{equation}
In particular it is straightforward to show that any bare exchange kernel satisfies $f^\lambda_{x}=\lambda f_{x}$.

\subsection{RPA renormalization}
While the Dyson equation \eqref{eq:dyson} provides an exact representation of $\chi$ for a given kernel, the solution of the equation may
exhibit pathological behavior related to electronic instabilities\cite{furche2005fluctuation,KTC11,CHG14}. The simplest examples are for the low-density homogeneous electron gas and stretched diatomics, where Colonna \textit{et al.}\cite{CHG14} demonstrated that the exact-exchange kernel leads to a divergence in the density-density response function computed via Eq. \eqref{eq:dyson}. To avoid this problem,
an exact refactorization of Eq. \eqref{eq:dyson} was introduced by Bates and Furche in 2013\cite{BF13},
\begin{align}\label{eq:dyson-rpar}
\chi_\lambda(\omega) & = \chi^{RPA}_\lambda(\omega) 
                    + \chi^{RPA}_\lambda(\omega) f_{xc}^{\lambda}(\omega) \chi_\lambda(\omega) \, ,\\
             & = \left[\chi^{-1}_{\lambda,RPA}(\omega) - f^{\lambda}_{xc}(\omega) \right]^{-1} \, .
\end{align}
The series expansion of $\chi(\omega)$ in powers of $\chi_{KS}(\omega)$ generates an unscreened perturbation theory that is equivalent to G\"{o}rling-Levy Perturbation theory\cite{GL93}. Since the bare Kohn-Sham orbital energy differences appear in the denominator of the non-interacting response function, the unscreened perturbation series diverges for small-gap or metallic systems.\cite{manybodybook} This divergence can be eliminated by expanding in powers of the RPA response function, 
\begin{align}
\chi^{RPA}(\omega) = \left[ \chi^{-1}_{KS}(\omega) - v_c \right]^{-1} \, ,
\end{align}
which leads to the following series
\begin{align}\label{eq:rpar-series}
\chi_\lambda(\omega) \approx 
\chi^{RPA}_\lambda(\omega) 
+ \chi^{RPA}_\lambda(\omega) f_{xc}^\lambda(\omega) \chi^{RPA}_\lambda(\omega) + \ldots
\end{align} 
In addition to eliminating the divergences related to the non-interacting response function, this expansion also eliminates the electronic instabilities resulting from the kernel since inversions are never needed directly involving the xc-kernel.\cite{BF13,CHG14,BSR17}

The decomposition in Eq. \eqref{eq:dyson-rpar} also naturally leads to a simple partition of the correlation energy into two pieces
\begin{align}\label{eq:ec-rpar}
E_c[f_{xc}] = E_c^{RPA} + \Delta E_c^{bRPA}[f_{xc}].
\end{align} 
The beyond-RPA (bRPA) piece incorporates all of the terms in Eq. \eqref{eq:rpar-series} beyond the ``bare'' RPA response function, which can be collected and exactly expressed as\cite{BSR17}
\begin{align}\label{eq:ec-brpar}
\Delta E_c^{bRPA} =-\int_0^1 d\lambda\int_0^\infty \frac{d\omega}{2\pi}\llangle v_c \chi^{RPA}_\lambda(\omega) f_{xc}^\lambda(\omega) \chi_\lambda(\omega)\rrangle.
\end{align} 
By truncating $\chi_\lambda$ to a low-order in $\chi_{\lambda}^{RPA}$, one hopes to faithfully reproduce the infinite-order correlation energy while avoiding the need to invert a function that directly contains $f_{xc}$. We stress that this approximation scheme can never exceed the accuracy of the infinite-order approach for energy differences and material properties, but it does guarantee the stability of the scheme to compute the correlation energy.\cite{BLR16} Furthermore, this division naturally separates the long-range and short-ranged contributions to the correlation energy, enabling approximations for $\Delta E_c^{bRPA}$ to be added directly on top of the already robust random phase approximation.

The first-order approximation derived from RPA renormalization, RPAr1, recovers a significant part ($\sim$90\%) of the total bRPA correlation energy for a given kernel\cite{BLR16,BSR17}
\begin{align}\label{eq:ec-rpar1}
\Delta E_c^{RPAr1} =-\int_0^1 d\lambda&\int_0^\infty \frac{d\omega}{2\pi}\llangle v_c \chi^{RPA}_\lambda(\omega) f_{xc}^\lambda(\omega) \chi^{RPA}_\lambda(\omega)\rrangle.
\end{align} 
This approximation has several key features: it recovers the exact, second-order correlation energy given the exact-exchange kernel, the coupling strength integral can be performed analytically for exchange-like kernels 
leading to efficient implementations\cite{BF13,BLR16,CAF18}, and it reduces properly to RPA for stretched bonds unlike other second-order schemes such as
SOSEX\cite{HS10}. In fact, within the adiabatic-connection framework, SOSEX can be obtained directly from RPAr1\cite{BF13,BLR16,BSR17} through the replacement of one $\chi^{RPA}$ with $\chi_{KS}$ in Eq.~\eqref{eq:ec-rpar1}
\begin{align}\label{eq:ec-acsosex}
\Delta E_c^{ACSOSEX} &=-\int_0^1 d\lambda \int_0^\infty \frac{d\omega}{2\pi} 
        \left\llangle v_c \chi^{RPA}_\lambda(\omega) f_{xc}^\lambda(\omega) \chi_{KS}(\omega)
        \right\rrangle.
\end{align} 
This approximation was shown to be less consistent than RPAr1 due to the reintroduction of the KS response function for molecular energy differences\cite{BF13} and structural properties of simple solids\cite{BLR16}.

To recover the remaining $\sim$10\% of the bRPA correlation energy, corrections beyond RPAr1 to the response function can be systematically added
order-by-order until convergence to Eq. \eqref{eq:dyson}. Rather than compute these terms exactly, a simple approximation can be introduced to eliminate the coupling-strength integration and utilize information from second-order to estimate third and higher order terms in the RPA renormalized expansion. This approximation method was termed the Higher-Order Terms (HOT) approximation\cite{BSS18} and is obtained through a rescaling of the second-order RPAr correction at $\lambda=1$. The HOT approximation usually reproduces the total correlation energy to within 1-2\%, and, consequently, accurately reproduces the performance of a given kernel for chemical or physical properties of molecules and materials.

\subsection{xc-kernels from the HEG}

Although Eq. \eqref{eq:f_xc} provides a definition of the kernel $f_{xc}$, the absence of an exact expression for the exchange-correlation potential $v_{xc}$ requires that an approximate form of $f_{xc}$ must be used in practical calculations. The homogeneous electron gas (HEG) provides a valuable testing ground for approximations of $f_{xc}$ and allows the kernel's limiting behaviour to be studied. The analogue of Eq. \eqref{eq:dyson} for the HEG is
\begin{align}
    \chi^\mathrm{HEG}(q,\omega)=&\chi_{0}(q,\omega)+\chi_{0}(q,\omega) \Big[v_c(q) + f^\mathrm{HEG}_{xc}(q,\omega)\Big]\chi^\mathrm{HEG}(q,\omega) \nonumber,
\end{align}
where the dependence on the wavevector $q$ has now been made explicit. $\chi_{0}$ is the textbook Lindhard function\cite{Inksonbook}, which coincides with $\chi_{KS}$ when Eq. \eqref{eq:dyson} is applied to the HEG\cite{Farid1993}. A quantity commonly found in the HEG literature is the local field factor $G(q,\omega)$, which is closely related to $f^\mathrm{HEG}_{xc}$ as  $ f^\mathrm{HEG}_{xc}(q,\omega) = -v_c (q)G(q,\omega)$.

Theoretical work on $G$ (and thus $f^\mathrm{HEG}_{xc}$) can be traced at least as far back as Hubbard (see Sec.\ III C of Ref. \cite{Ichimaru1982}
for a review), and exact limits have been derived for a number of cases. First, the long wavelength and static limit ($q \rightarrow 0$, $\omega =0 $) actually corresponds to the adiabatic local density approximation (ALDA) commonly employed in TDDFT,
\begin{align}\label{eq:limitALDA}
    f^\mathrm{HEG}_{xc}(q\rightarrow0,\omega=0)= f^\mathrm{ALDA}_{xc} \equiv - \frac{4\pi A}{k_F^2}
\end{align}
where
\begin{align}\label{eq:limitA}
    A = \frac{1}{4} - \frac{k_F^2}{4\pi} \frac{d^2(nE_C)}{dn^2},
\end{align}
$k_F = (3\pi^2n)^{1/3}$ is the Fermi wavevector for the HEG of density $n$, and $E_C$ is the correlation energy per electron. The two terms in Eq. \eqref{eq:limitA} correspond to exchange and correlation contributions. Eq. \eqref{eq:limitALDA} is intuitive in stating that the ALDA should be exact in describing the HEG response to a uniform, static perturbation\cite{Constantin2011}, and is more formally derived from the 
compressibility sum rule\cite{Ichimaru1982}. Next, the short wavelength and static limit has the form\cite{Holas1987,Toulouse2005}
\begin{align}\label{eq:shortwave}
    f^\mathrm{HEG}_{xc}(q\rightarrow\infty,\omega=0)=  - \frac{4\pi B}{q^2} - \frac{4\pi C}{k_F^2},
\end{align}
while the long wavelength, high frequency limit has the form\cite{Gross1985}
\begin{align}\label{eq:highfreq}
    f^\mathrm{HEG}_{xc}(q=0,\omega\rightarrow\infty)=  - \frac{4\pi D}{k_F^2}.
\end{align}
The parameters $A$, $B$, $C$ and $D$ depend on the HEG density, which in turn can be written in terms of the Fermi wavevector or Wigner radius $r_s = (3/4\pi n)^{1/3}$. Practically, $A$, $C$ and $D$ can be obtained from a parameterization of the correlation energy $E_C$ for a HEG of density $n$, wherase $B$ requires additional knowledge of the momentum distributio of the HEG\cite{Ortiz1994}.

For intermediate $q$ values it is necessary to turn to diffusion Monte Carlo calculations. The study of Ref. \cite{Moroni1995} investigated the $q$-dependence of the static kernel $f^\mathrm{HEG}_{xc}(q,\omega=0)$ for a range of densities. A key conclusion of that work was that for wavevectors $q\leq 2k_F$,  $f^\mathrm{HEG}_{xc}(q,\omega=0)$ remains close to its $q=0$ value,  (=$f^\mathrm{ALDA}_{xc}$, Eq.  \eqref{eq:limitALDA}), while for $q>2k_F$,  the kernel can be reasonably well described by the short wavelength limit (Eq. \eqref{eq:shortwave}).

In the context of approximations to $f^\mathrm{HEG}_{xc}$, it is worth stressing a point discussed in Ref. \cite{Farid1993}: there is no particular reason why approximate, frequency-independent kernels should
display the same limiting behaviour as the exact, frequency-dependent kernel evaluated at $\omega = 0$. Indeed, having a frequency-independent kernel which is finite at large $q$ (obeying Eq. \eqref{eq:shortwave}) will in fact lead to a pair-distribution function which is singular at the origin\cite{Kimball1973,Singwi1970,Furche2005}.

\begin{figure}[tb]
\begin{center}
 \includegraphics[scale=1.0]{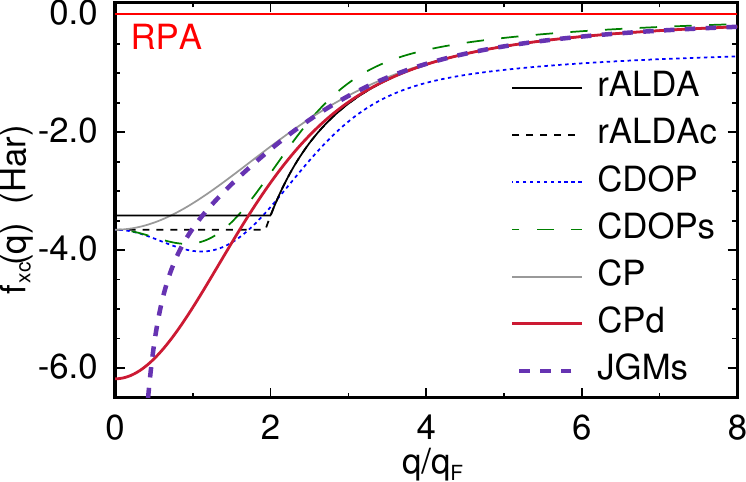}
 \caption{Various approximations for the exchange-correlation kernels applied
to the homogeneous electron gas as a function of wavevector, evaluated at a density corresponding to $r_s = 2.0$ bohr radii. The dynamical CPd kernel was evaluated at $\omega$ = 2 Hartrees and the JGMs kernel was evaluated at a band gap of 3.4~eV. The trivial RPA case  ($f^\mathrm{HEG}_{xc}=0$) is also shown.}
\label{fig:kernels}
\end{center}
\end{figure} 

In Fig.~\ref{fig:kernels} we show some approximate forms for $f^\mathrm{HEG}_{xc}$, which have been proposed in the literature\cite{Olsen2012,Corradini1998,Lu2014,Constantin2007}. The ``rALDA'' kernel will be discussed in some detail in the following sections. Briefly describing the other kernels, ``CDOP'' refers to the frequency-independent 
kernel proposed by Corradini, Del Sole, Onida and Palummo\cite{Corradini1998}, which has the same limiting behaviour as the exact static kernel (Eqs. \eqref{eq:limitALDA} and \eqref{eq:shortwave}). ``CDOPs'' refers to the kernel introduced in Ref.~\cite{Lu2014}, which modifies CDOP so that it vanishes at large $q$. ``CPd'' refers to the dynamical kernel proposed by Constantin and Pitarke\cite{Constantin2007}, 
which satisfies the long wavelength static and high frequency limits (Eqs. \eqref{eq:limitALDA} and \eqref{eq:highfreq}). The frequency-independent ``CP'' kernel corresponds to the CPd kernel at $\omega=0$.

Although the HEG carries the advantage of being a very well-studied system, it is worth remembering that fundamentally it is metallic. The exchange-correlation kernel of a periodic insulator is known to display different limiting behaviour to that of a metal, diverging as $1/q^2$ in the $q\rightarrow0$ limit\cite{Aulbur1996,Ghosez1997}. This aspect is especially important in TDDFT calculations of optical spectra including excitonic effects\cite{Trevisanutto2013,Botti2004,Sharma2011}. This consideration led to the development of the frequency-independent jellium-with-gap model kernel, which has the $1/q^2$ divergence\cite{Trevisanutto2013}. The slightly simpler ``JGMs'' kernel shown in Fig. \ref{fig:kernels} is described in Ref. \cite{Patrick2015}. Here the band gap $E_g$ enters parametrically. In the limit $E_g\rightarrow \infty$ the correlation energy disappears, while  $E_g\rightarrow 0$ the metallic CP kernel is recovered.

Ref. \cite{Patrick2015} provides a more thorough discussion of all of the kernels shown in Fig. \ref{fig:kernels}, including the expressions used to evaluate them and their forms in real space. In the current work we focus our attention on the renormalized adiabatic kernels rALDA and rAPBE, although a comparison of all the different xc-kernels for the structural parameters of solids will be presented in Sec. \ref{sec:structure}.

\subsection{The renormalized adiabatic LDA kernel}
Ideally one should aim at obtaining a general approximation to $f_{xc}$ that can reproduce various physical quantities such as optical absorption spectra and ground state electronic correlation energies. However, finding good approximations for $f_{xc}$ is highly challenging and it is often necessary to limit the approximation to a given application. As mentioned previously, it is crucial to use a form of $f_{xc}$ that has the correct $1/q^2$ behavior in the long wavelength limit in order to capture excitonic effects in absorption spectra. On the other hand, ground state correlation energies involve $q$-space integrals making it extremely important to obtain a good approximation at large values of $q$, whereas the long wavelength limit is less important. In the following we will focus on obtaining an approximation that provides accurate ground state correlation energies. 

The correlation energy per electron is directly related to the integral of the coupling constant averaged correlation hole $\bar{n}_c(r)$\cite{Lein2000}
\begin{equation}\label{eq:hem}
 E_c=2\pi\int_0^\infty dr r\bar{n}_c(r)=\frac{1}{\pi}\int_0^\infty dq \bar{n}_c(q).
\end{equation}
where $\bar n_c(q)$ is the Fourier transform of $\bar n_c(r)$.  A parametrization of the exact $n_c(q)$ has been provided by Perdew and Wang\cite{perdew_wang} based on quantum Monte Carlo simulations of the homogeneous electron gas at various densities. Approximations for $n_c(q)$ can be obtained from Eqs. \eqref{eq:E_c_x} and \eqref{eq:dyson} using the Lindhard function for $\chi_0(q,\omega)$. 

\begin{figure}[tb]
	\includegraphics[width=4.25 cm]{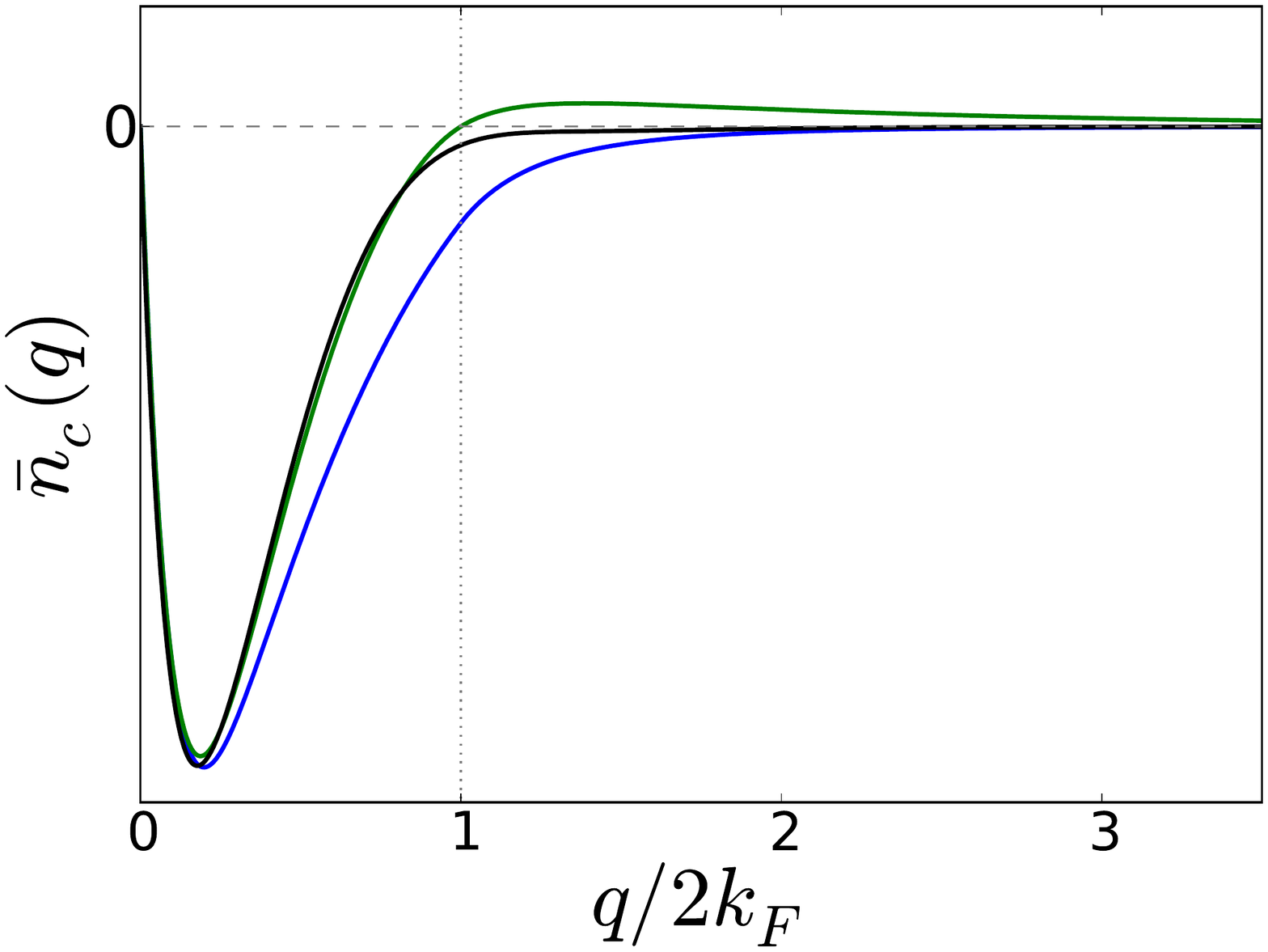} 
    \includegraphics[width=4.25 cm]{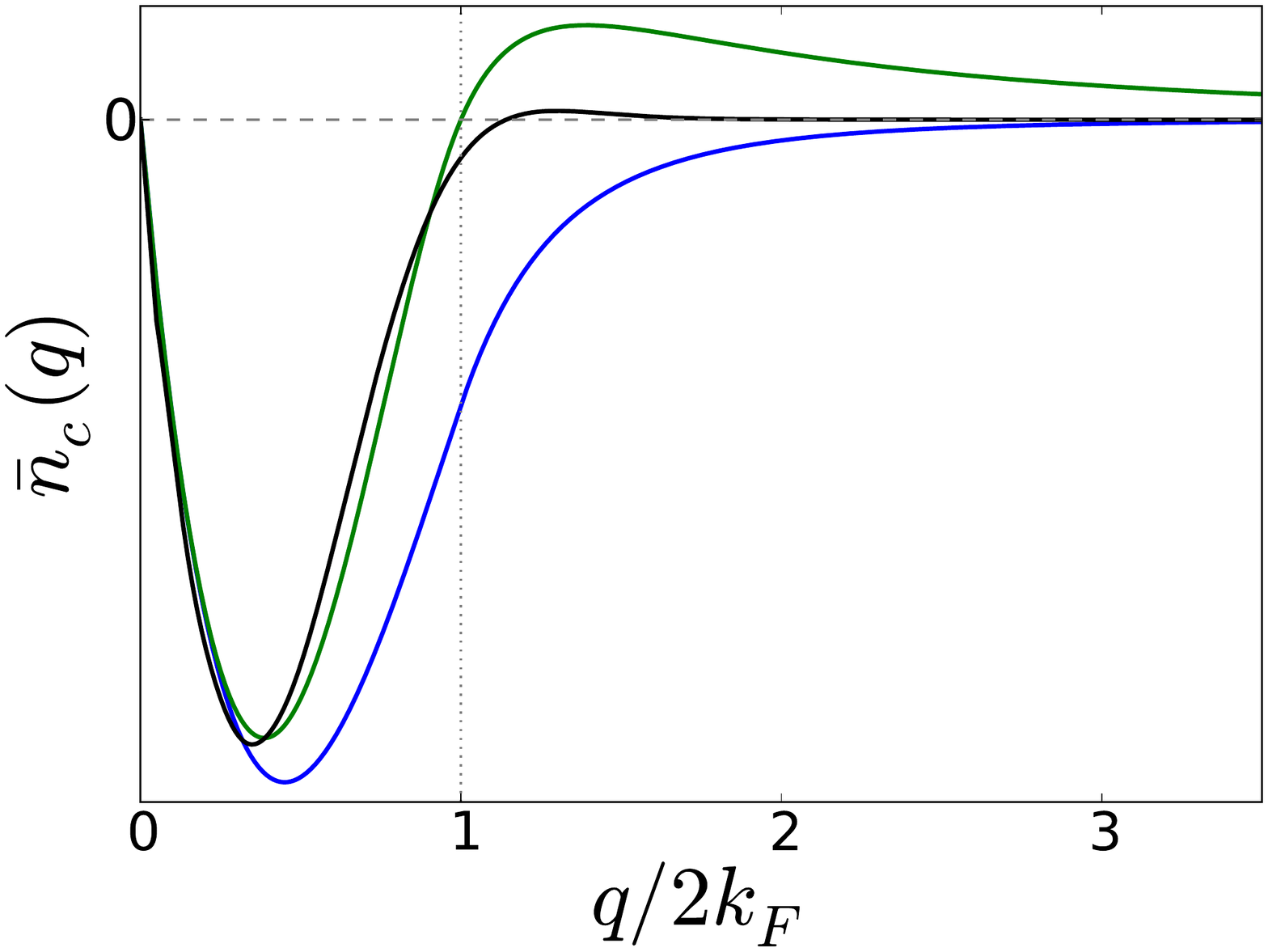}
\caption{Correlation hole of the homogeneous electron gas  in $q$-space at $r_s=1$ (left) and $r_s=10$ (right).}
\label{fig:g_q}
\end{figure}
The correlation hole in $q$-space is shown in Fig. \ref{fig:g_q} calculated with RPA and ALDA for two different densities. Compared to the exact parametrization it is clear that RPA severely overestimates the magnitude of the correlation hole and the RPA will predict a correlation energy that is $\sim0.5$ eV too low per electron for a wide range of densities. The ALDA on the other hand straddles the exact parametrization for a wide range of $q$-values but decays too slowly at large $q$ compared to the exact results. This is a consequence of the locality of the approximation, which translates into an independence of $q$. At large $q$ the xc-kernel will thus dominate the Coulomb kernel and fail to reproduce the exact limit \eqref{eq:shortwave}. Since the total energy involves a $q$-space integral over all space the slow decay of the correlation hole introduces significant errors and overestimates the correlation energy by $\sim0.3$ eV per electron.

\begin{figure}[tb]
	\includegraphics[width=8 cm]{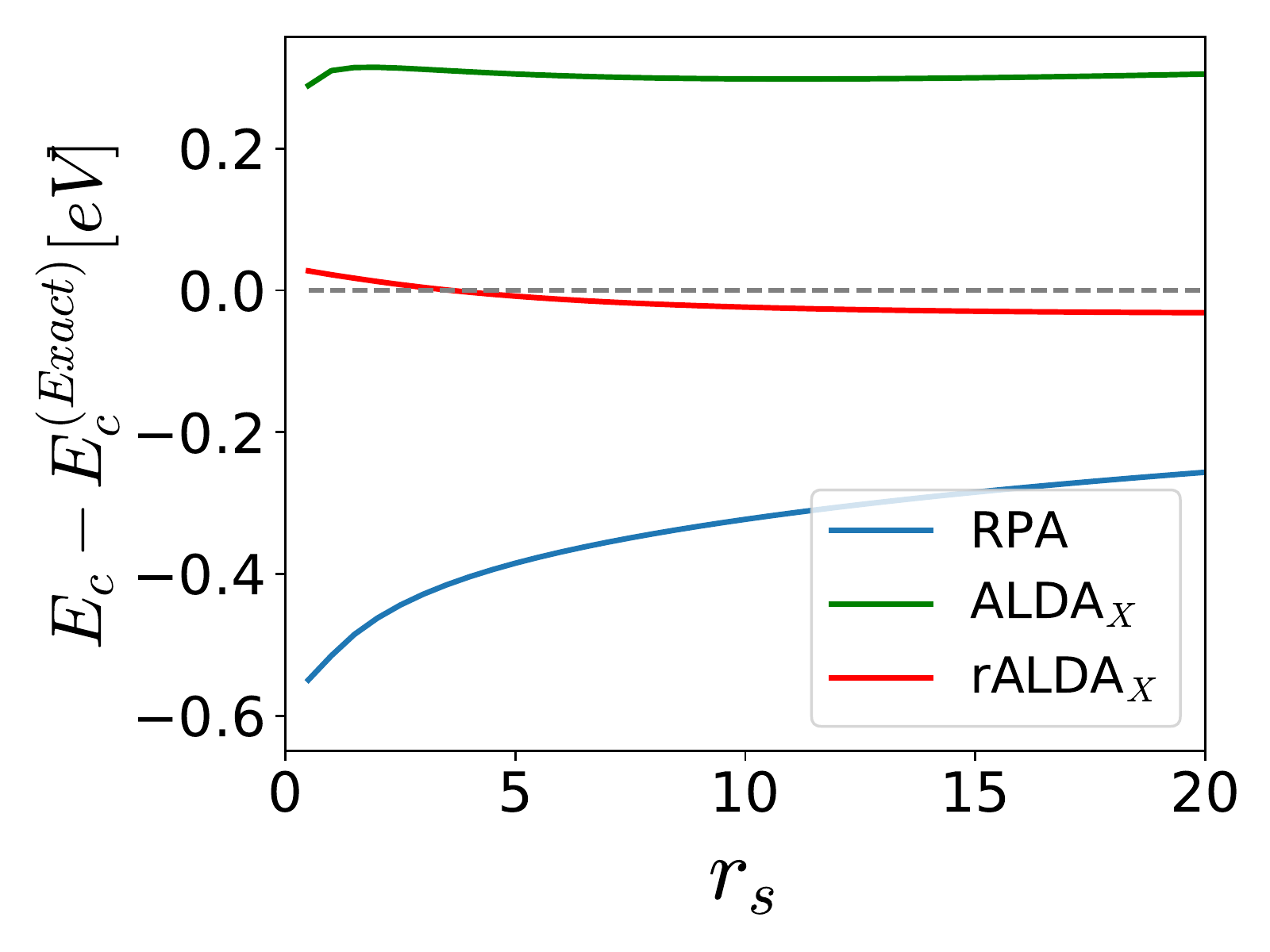} 
\caption{Correlation energy per electron of the homogeneous electron gas evaluated with the RPA, ALDA$_x$ and rALDA$_x$.}
\label{fig:e_heg}
\end{figure}

The ALDA$_x$ kernel provides a good approximation to the exact one for both low $r_s=1$ and high $r_s=10$ densities for $q<2k_F$, where the correlation hole has a zero point in $q$-space. However, for $q>2k_F$ the exact correlation hole largely vanishes and we expect to obtain a better approximation for the correlation energy if we simply truncate the $q$-integration at $2k_F$ when evaluating \eqref{eq:hem} using the ALDA$_x$ approximation. We will refer to this scheme as renormalized ALDA$_x$ (rALDA), since the truncation preserves the integral of the correlation hole in real space.  The correlation energy per electron evaluated in this scheme is shown in Fig. \ref{fig:e_heg} obtained with RPA, ALDA$_x$, and rALDA. Evidently, the errors in the correlation energy obtained with rALDA are much smaller than both RPA and ALDA$_x$.

For the homogeneous electron gas, the truncation is equivalent to using the Hartree-exchange-correlation kernel
\begin{align}
f_{Hxc}^{rALDA}[n](q)=\theta\Big(2k_F-q\Big)f_{Hx}^{ALDA}[n].
\end{align}
Fourier transforming this expression yields
\begin{align}\label{eq:rALDA}
f_{Hxc}^{rALDA}[n](r)&=\widetilde f_{xc}^{rALDA}[n](r)+v^r[n](r),\\ 
\widetilde f_{xc}^{rALDA}[n](r)&=\frac{f^{ALDA}_{x}[n]}{2\pi^2r^3}\Big[\sin(2k_Fr)-2k_Fr\cos(2k_Fr)\Big],\notag\\
v^r[n](r)&=\frac{1}{r}\frac{2}{\pi}\int_0^{2k_Fr}\frac{\sin x}{x}dx.\notag
\end{align}
Since $k_F$ is related to the density, one can attempt to generalize this scheme to inhomogeneous systems. We then take $r\rightarrow|\mathbf{r}-\mathbf{r}'|$ and $k_F\rightarrow(3\pi^2 n(\mathbf{r},\mathbf{r}'))^{1/3}$, but there is no unique way to define the two-point density $n(\mathbf{r},\mathbf{r}')$. A natural choice is to take\cite{Lein2000} $n(\mathbf{r},\mathbf{r}')=(n(\mathbf{r})+n(\mathbf{r}'))/2$, but other choices are possible, which will be discussed in Sec. \ref{sec:implementation}.

The simple truncation procedure has thus led to a non-local rALDA kernel that does not contain any free parameters and significantly improves correlation energies for homogeneous systems. We have split the Hartree-exchange-correlation kernel into a renormalized exchange-correlation part $f_{xc}^{rALDA}[n]$ and a renormalized Coulomb part $v^r[n]$. However, both terms depend on the density and contain exchange-correlation effects. The $\widetilde f_{xc}^{rALDA}$ part can be regarded as an ALDA$_x$ kernel where the delta function has acquired a density dependent broadening, whereas $v^r$ is the Coulomb interaction reduced by a density and distance dependent factor that approaches unity for large densities or distances. In fact, at large separation $f_{Hxc}^{rALDA}$ reduces to the pure Coulomb kernel and it is expected to retain the accurate description of long range van der Waals interactions within RPA. For example, in a jellium with $r_s=2.0$ two points separated by 5 {\AA} gives a renormalized Coulomb interaction  $v^r[r_s=2](r)=0.97v(r)$ and the magnitude is a factor of 30 times larger than $\widetilde f_{xc}^{rALDA}$. Interestingly, both $v^r$ and $\widetilde f_{xc}^{rALDA}$ becomes finite at the origin giving
\begin{align}
v^r[n](r\rightarrow0)&=\frac{4k_F}{\pi}-\frac{8k_F^3r^2}{9\pi},\label{kernel_origin}\\
\widetilde{f}_{xc}^{rALDA}[n](r\rightarrow0)&=\Big[\frac{4k_F^3}{3\pi^2}-\frac{32k_F^5r^2}{15\pi^2}\Big]f^{ALDA}_{x}[n],
\end{align}
which implies $f^{rALDA}_{Hxc}[n](r=0)=0$. This property is related to the fact that the position weighted correlation hole entering the first integral in Eq. \eqref{eq:hem} vanishes at the origin\cite{Olsen2013a} and is highly convenient for numerical real-space evaluation of the kernel.

It is often more convenient to separate the Hxc-kernel into the exact Coulomb kernel and an xc-kernel and one is then led to define
\begin{align}\label{eq:rALDA_exchange}
f_{x}^{rALDA}[n](r)&=\widetilde{f}_{x}^{rALDA}[n](r)+v^r[n](r)-v(r).
\end{align}
This expression is typically more useful for applications to periodic systems since  $v^r[n](r)-v(r)$ is short ranged ([$v^r[n](r)-v(r)]\rightarrow\sin(2k_Fr)/r$ for $r\rightarrow\infty$) whereas both $v^r[n](r)$ and $v(r)$ are long ranged.

\subsubsection{Generalized truncation scheme}\label{sec:gen_renorm_scheme}
The truncation scheme defined above is easily generalized to any adiabatic semi-local local kernel. The correlation hole of the homogeneous electron gas calculated with ALDA$_x$ has a zero point exactly at $2k_F$. This is not true in general, but for any adiabatic kernel the correlation hole becomes zero at the point where the Hartree-exchange-correlation kernel vanishes. This leads to the zero-point wavevector
\begin{align}\label{eq:q_0}
q_0[n]=\sqrt{\frac{-4\pi}{f_{xc}^{A}[n]}},
\end{align}
where $f_{xc}^{A}[n]$ is the spatial Fourier transform of the adiabatic kernel
\begin{align}
f_{xc}^A[n](\mathbf{r}-\mathbf{r}')=\frac{\delta v_{xc}(\mathbf{r})}{\delta n(\mathbf{r}')}\delta(\mathbf{r}-\mathbf{r}'),
\end{align}
corresponding to a particular semi-local approximation. Renormalized kernels for any semi-local approximation for the exchange-correlation functional can then be defined by replacing to $2k_F$ by $q_0$ in Eq. \eqref{eq:rALDA} and the kernel is again generalized to inhomogeneous systems by taking $r\rightarrow|\mathbf{r}-\mathbf{r}'|$ and $n\rightarrow n(\mathbf{r},\mathbf{r}')$ in addition to a scheme that defines $n(\mathbf{r},\mathbf{r}')$ in terms of $n(\mathbf{r})$. For generalized gradient corrected functionals, $q_0$ will depend on the gradient of the density as well, which may lead to positive values of $f_{xc}^A$ at certain points. At those points we set $q_0[n](\mathbf{r},\mathbf{r}')=0$ in order to maintain a well defined kernel. Below we will only consider rALDA and rAPBE.

\subsection{Spin}\label{sec:spin}
The inclusion of spin degrees of freedom in RPA is almost trivial since the correlation energy involves the sum over all spin components $\chi_{\sigma\sigma'}$, which is obtanined by the simple substitution $\chi^0\rightarrow \chi^0_{\uparrow\uparrow}+\chi^0_{\downarrow\downarrow}$ in Eq. \eqref{eq:dyson}. This is due to the fact that $F_{Hxc}$ is independent of spin in RPA. In general, however, it is not straightforward to generalize a kernell for spin-paired systems to the spin-polarized case.

In the case of exchange one can resort to the spin dependence of the exchange energy. In particular one has
\begin{align}\label{Exs}
E_x[n_\uparrow,n_\downarrow]=\frac{E_x[2n_\uparrow]+E_x[2n_\downarrow]}{2},
\end{align}
which yields
\begin{align}\label{eq:fxs}
f_{x,\sigma\sigma'}[n_\uparrow,n_\downarrow]=2f_x[2n_\sigma]\delta_{\sigma\sigma'}.
\end{align}
It is possible to enforce this condition on the rALDA kernel as well, but we have found that it makes the correlation energy difficult to converge. The reason is that the off-diagonal (in spin) componenets of the Hxc-kernel involves a bare Coulomb interaction, whereas the diagonal components lack a long-range cancellation between $v(r)$ and $v^r[n]$.

This failure is clearly a limitation of the rALDA scheme and an additional approximation is required to maintain the the accuracy of rALDA for spin-polarized systems. To this end we start with the Dyson equation with explicit spin dependence
\begin{align}\label{dyson1}
\chi_{\sigma\sigma'}=\chi^{KS}_{\sigma}\delta_{\sigma\sigma'}+\sum_{\sigma''}\chi^{KS}_{\sigma}f^{Hxc}_{\sigma\sigma''}[n_\uparrow,n_\downarrow]\chi_{\sigma''\sigma'},
\end{align}
where it was used that $\chi^{KS}$ is diagonal in spin. For the spin-paired case one has that
\begin{align}\label{eq:dyson_spin_condition}
\frac{1}{4}\sum_{\sigma\sigma'}f^{xc}_{\sigma\sigma'}[n/2,n/2]=f_x[n],
\end{align}
which will always hold if Eq. \eqref{eq:fxs} if Eq. \eqref{eq:fxs} is satisfied. To reintroduce a balanced expression for the renormalized kernel in each spin component we relax Eq. Eq. \eqref{eq:fxs} and use 
\begin{align}\label{eq:rALDA_spin}
f^{rALDA}_{xc,\sigma\sigma'}[n_\uparrow,n_\downarrow]=2f^{rALDA}_{xc}[n]\delta_{\sigma\sigma'}+v^r[n]-v,
\end{align}
where $n=n_\sigma+n_{\sigma'}$. Eq. \eqref{eq:dyson_spin_condition} is now satisfied, but Eq. \eqref{eq:fxs} is not. This choice is not unique though and another choice is comprised by $f^{rALDA}_{x,\sigma\sigma'}=2f^{rALDA}_x[n_\sigma+n_{\sigma'}]\delta_{\sigma\sigma'}+v^r[n_\sigma+n_{\sigma'}]-v$, which was used in Ref. \cite{Olsen2012}. However, Eq. \eqref{eq:rALDA_spin} appears to yield better results for atomization energies where spin-polarized isolated atoms are used as a reference.

\subsection{Hedin's equations and vertex corrections}\label{sec:hedin}
So far we have discussed the use of xc-kernels in the context of the ACFDT formula for the ground state correlation energy. However, it is possible, and in fact quite effective, to apply the same xc-kernels to describe the effect of vertex corrections in the electron self-energy. In 1965 Lars Hedin introduced a set of coupled equations relating the single-particle Green's function $G$, the electron self-energy, $\Sigma$, to the polarization, $P$, the screened electron-electron interaction, $W$, and the 3-point vertex function $\Gamma$\cite{hedin1965new},
\begin{eqnarray}
G(1,2)&=&G_H(12)+\int d(34)G_H(1,3)\Sigma(3,4)G(4,2)\\
\Sigma(1,2)&=&i\int d(34)G(1,3)\Gamma(3,2,4) W(4,1^+)\\
W(1,2)&=&v(1,2) + \int d(34)W(1,3)P(3,4)v(4,2)\\
P(1,2)&=&-i\int d(3,4) G(1,3)G(4,1^+)\Gamma(3,4,2)\\
\Gamma(1,2,3)&=&\delta(1,2)\delta(1,3)\\
&+&\int d(4567)\frac{\delta \Sigma(1,2)}{\delta G(4,5)}G(4,6)G(7,5)\Gamma(6,7,3)\notag,
\end{eqnarray}
where we employed the notation $(1)=(\mathbf{r}_1,t_1,\sigma_1)$ and $G_H$ is the Hartree Green's function. The well known and widely used GW approximation is obtained by iterating Hedin's equations once starting from $\Sigma=0$, i.e. the Hartree approximation. This produces the trivial vertex function $\Gamma=\delta(1,2)\delta(1,3)$, which corresponds to the time-dependent Hartree approximation for the polarization $P$, which is the approximation refeered to as RPA in the present review. There are basically two issues with this approach. First of all, it starts from $G_H$, which is known to be a poor approximation. Secondly, it neglects vertex corrections completely. In practice, the latter issue is rarely dealt with because of the complex nature of $\Gamma$, while the first is overcome by following a "best $G$, best $W$" philosophy\cite{hybertsen_first-principles_1985}. Within the popular G$_0$W$_0$ method, one evaluates the self-energy from a non-interacting $G_0$ obtained from a DFT calculation while $W$ is obtained within RPA using the polarization $P_0=G_0G_0$. Today, the G$_0$W$_0$ method remains the state-of-the-art for calculation of QP band structures of inorganic solids\cite{van2006quasiparticle,shishkin2006implementation,scherpelz2016implementation} and nano-structures including two-dimensional materials\cite{rasmussen2016efficient,qiu2016screening,haastrup2018computational}.

Another question related to Hedin's equation is the role of self-consistency. In principle, the five equations should be solved self-consistently. However, while self-consistency improves the description of energy levels in molecules\cite{rostgaard_fully_2010,blase_first-principles_2011} and is essential for systems out of equilibrium\cite{baym,thygesen2008conserving,strange2011self,myohanen2009kadanoff}, it does not in general improve the the band structure and spectral functions of solids when vertex corrections are neglected\cite{shishkin2007self,von1996self}. The role of self-consistency will not be further discussed in this review where we instead concentrate on the problems of vertex corrections.    

Rather than starting the iterative solution of Hedin's equations with $\Sigma=0$ (which leads to the GW approximation at first iteration), it is of course possible to start with a local approximation, $\Sigma^0(1,2)=\delta(1,2)v_{xc}(1)$. As shown by Del Sole \textit{et al.}\cite{delsole} this leads to a self-energy of the form 
\begin{equation}\label{eq.gw}
\Sigma(1,2) = i G(1,2) \widetilde W(1,2),
\end{equation} 
where 
\begin{equation}\label{eq.tildeW}
\widetilde W = v [1-P_0(v+f_{xc})]^{-1}
\end{equation} 
and 
\begin{equation}\label{eq.fxc_vxc}
f_{xc}(1,2)=\delta v_{xc}(1)/\delta n(2)
\end{equation}
is the adiabatic xc-kernel. By inspection it becomes clear that $\widetilde W(1,2)$ is the screened \textit{effective} potential generated at point 2 by a charge at point 1. This potential consists of the bare Coulomb potential plus the induced Hartree and xc-potential. Consequently, it represents the potential felt by an electron. In contrast, the potential felt by a classical test charge is the bare potential screened only by the induced Hartree potential:       
\begin{equation}\label{eq.RPA_W}
W = v + v[1-P_0(v+f_{xc})]^{-1}P_0 v
\end{equation} 
In Eq. (\ref{eq.gw}) the replacement of $\widetilde W$ by $W$ thus corresponds to including the vertex in the polarisability, $P$, but neglecting it in the self-energy. In the following we refer to these two alternative schemes as G$_0$W$_0\Gamma_0$ and G$_0$W$_0$P$_0$, respectively. As usual the subscripts indicate that the quantities are evaluated non self-consistently starting from DFT. In addition to the fact that the vertex correction accounts for the change in the xc-potential and therefore should be more accurate, an attractive feature of the G$_0$W$_0\Gamma_0$ scheme is that DFT becomes the consistent starting point for non-selfconsistent calculations when the relation \eqref{eq.fxc_vxc} is satisfied. This is in stark contrast to G$_0$W$_0$ for which the Hartree approximation is the consistent starting point.

In Sec. \ref{sec.vertex_results} we show that when the rALDA xc-kernel is used to include vertex corrections through Eq. (\ref{eq.tildeW}) the improved description of short-range correlations lead to a significant up-shift of QP energies by 0.3-0.5 eV in agreement with experiments\cite{schmidt2017simple}. Since both occupied and unoccupied states are shifted up, the band gap is not affected as much but a small increase is generally observed again in agreement with experiments.

\section{Implementation}\label{sec:implementation}
\subsection{Evaluating non-local kernels for inhomogeneous densities}\label{sec:symmetrization}
Kernels like the rALDA which were derived from the HEG (a uniform system) have the form $f^m_{xc}[n](q,\omega)$ in
reciprocal space or $f^m_{xc}[n](|\mathbf{r}-\mathbf{r'}|,\omega)$ in real space. As mentioned in Sec. \ref{sec:gen_renorm_scheme}, this
nonlocality $|\mathbf{r}-\mathbf{r'}|$ leads to a question regarding the treatment of the density argument
when calculating the correlation energy of inhomogeneous systems [$n = n(\mathbf{r})$]. To illustrate this point more clearly, we consider the plane wave representation of the kernel,
\begin{align}\label{eq:FT}
f_{xc}^{\mathbf{G}\mathbf{G'}}(\mathbf{q},\omega) = \frac{1}{V} 
\int_V d\mathbf{r}&\int_V d\mathbf{r'} e^{-i(\mathbf{q}+\mathbf{G})\cdot\mathbf{r}}f_{xc}(\mathbf{r},\mathbf{r'},\omega) e^{i(\mathbf{q}+\mathbf{G'})\cdot\mathbf{r'}}.\notag
\end{align}
Here $V$ is the volume of the crystal, $\mathbf{G}$ and $\mathbf{G'}$ are reciprocal lattice vectors and $\mathbf{q}$ lies within the first Brillouin zone. In the case that the system under investigation is homogeneous [$n(\mathbf{r}) = n_0$] then the kernel is diagonal,
\begin{align}
f_{xc}^{\mathbf{G}\mathbf{G'}}(\mathbf{q},\omega) = \delta_{\mathbf{G}\mathbf{G'}}f^m_{xc}[n_0](|\mathbf{q}+\mathbf{G}|,\omega)
\end{align}
On the other hand, if the kernel is fully local (independent of $q$, e.g.\ the ALDA) it is natural to use the local density to evaluate the kernel, obtaining
\begin{align}
f_{xc}^{\mathbf{G}\mathbf{G'}}(\omega)=\frac{1}{\Omega}  \int_\Omega
d\mathbf{r} e^{-i(\mathbf{G}-\mathbf{G'})\cdot\mathbf{r}}f^m_{xc}[n(\mathbf{r})](\omega)
\end{align}
where $\Omega$ is the unit cell volume. However, for the general case of a inhomogeneous system and non-local kernel there is no unique way of constructing $f_{xc}[n](\mathbf{r},\mathbf{r'})$ from the knowledge of $f_{xc}[n](\mathbf{r})$ in a homogeneous system. The problem is that the density is a one-point function and it is not clear how to treat the dependence of $\mathbf{r}$ and $\mathbf{r'}$. One important constraint is that the resulting kernel must be symmetric in $\mathbf{r}$ and $\mathbf{r'}$\cite{Burkebook}, and in the following we assume the form
\begin{align}\label{eq:symkern}
f_{xc}(\mathbf{r},\mathbf{r'}) \rightarrow 
f^m_{xc}(\mathcal{S}[n],|\mathbf{r}-\mathbf{r'}|),
\end{align}
where $\mathcal{S}$ is a functional of the density symmetric in $\mathbf{r}$ and $\mathbf{r'}$ and we have restricted ourselves to frequency-independent kernels.

\subsubsection{Density symmetrization}
The density symmetrization scheme used in the rALDA/rAPBE calculations in Refs. \cite{Olsen2012,Olsen2013,Olsen2014a} employed a two-point average,
\begin{align}\label{eq:S}
\mathcal{S}[n] = [n(\mathbf{r}) + n(\mathbf{r'})]/2,
\end{align}
but more elaborate functionals are possible\cite{Garcia1996,Cuevas2012}. A kernel satisfying Eq. \eqref{eq:symkern} with a general two-point density is only invariant under simultaneous lattice translation in $\mathbf{r}$ and $\mathbf{r}'$. Its plane wave representation can then be written in the form
\begin{align}\label{f_GG}
f_{xc}^{\mathbf{G}\mathbf{G}'}(\mathbf{q})=\frac{1}{\Omega}\int_{\Omega}d\mathbf{r}\int_{\Omega}d\mathbf{r}'e^{-i\mathbf{G}\cdot\mathbf{r}}f(\mathbf{q};\mathbf{r},\mathbf{r}')e^{i\mathbf{G}'\cdot\mathbf{r}'}
\end{align}
where
\begin{align}\label{f_tilde}
f(\mathbf{q};\mathbf{r},\mathbf{r}')=\frac{1}{N}\sum_{i,j}e^{i\mathbf{q}\cdot\mathbf{R}_{ij}}e^{-i\mathbf{q}\cdot(\mathbf{r}-\mathbf{r}')}f^{xc}_{x}(\mathbf{r},\mathbf{r}'+\mathbf{R}_{ij}).
\end{align}
and $\mathbf{R}_{ij}=\mathbf{R}_i-\mathbf{R}_j$. $f(\mathbf{q};\mathbf{r},\mathbf{r}')$ is periodic in $\mathbf{r}$ and $\mathbf{r}'$ and Eq. \eqref{f_GG} must be converged by unit cell sampling, which should typically match the $k$-point sampling in periodic systems.

\subsubsection{Kernel symmetrization}
A second approach symmetrizes the kernel itself\cite{Lu2014}. Starting from a non-symmetric kernel, 
\begin{align}
f^{\mathrm{NS}}_{xc}(\mathbf{r},\mathbf{r'},\omega)
= f^m_{xc}[n(\mathbf{r})](|\mathbf{r}-\mathbf{r'}|,\omega)
\end{align}
and inserting into equation~\ref{eq:FT} gives
\begin{align}\label{eq:NS}
f^{\mathrm{NS},\mathbf{G}\mathbf{G'}}_{xc}(\mathbf{q},\omega)
= \frac{1}{\Omega}\int_\Omega 
d\mathbf{r} e^{-i(\mathbf{G}-\mathbf{G'})\cdot\mathbf{r}} 
f^m_{xc}[n(\mathbf{r})])(|\mathbf{q}+\mathbf{G}|,\omega).
\end{align}
It is now possible to obtain asymmetric kernel by taking the average $f^{\mathrm{NS},\mathbf{G}\mathbf{G'}}_{xc}(\mathbf{q},\omega)$ and its Hermitian conjugate,
\begin{align}
f^{\mathrm{S},\mathbf{G}\mathbf{G'}}_{xc}(\mathbf{q},\omega) = \frac{1}{2}\left(
f^{\mathrm{NS},\mathbf{G}\mathbf{G'}}_{xc}(\mathbf{q},\omega) + [f^{\mathrm{NS},\mathbf{G'}\mathbf{G}}_{xc}(\mathbf{q},\omega)]^*\right)
\end{align}
which can be seen equivalently as inserting the two-point 
average  $1/2[f^{\mathrm{NS}}_{xc}(\mathbf{r},\mathbf{r'},\omega) +
f^{\mathrm{NS}}_{xc}(\mathbf{r'},\mathbf{r},\omega)]$ into Eq. \eqref{eq:FT}\cite{Lu2014}. Compared to density symmetrization, Eq. \eqref{eq:NS} has the advantage that the integral is performed over one unit cell only, and that only the density has to be represented on a real space grid, while the kernel is defined by its plane wave representation.

\subsubsection{Wavevector symmetrization}
The third approach we consider is that employed in Ref. \cite{Trevisanutto2013}, which retains the computational advantages of the kernel symmetrization scheme. Here the wavevector  $|\mathbf{q}+\mathbf{G}|$ entering Eq. \eqref{eq:NS} is replaced by the symmetrized quantity  $\sqrt{|\mathbf{q}+\mathbf{G}||\mathbf{q}+\mathbf{G'}|}$, such that
\begin{align}\label{eq:wavesym}
f^{\mathbf{G}\mathbf{G'}}_{xc}(\mathbf{q},\omega)
&=\frac{1}{\Omega}\int_\Omega 
d\mathbf{r} e^{-i(\mathbf{G}-\mathbf{G'})\cdot\mathbf{r}} f^m_{xc}[n(\mathbf{r})](\sqrt{|\mathbf{q}+\mathbf{G}||\mathbf{q}+\mathbf{G'}|},\omega).
\end{align}
A kernel constructed using Eq. \eqref{eq:wavesym} will automatically satisfy the symmetry requirement of Eq. \eqref{eq:symkern}. Furthermore, for the specific case of kernels based on the jellium-with-gap model, the head and wings of the matrix in $\mathbf{G}$ and $\mathbf{G'}$ have the correct $1/q^2$ and $1/q$ divergences, respectively\cite{Trevisanutto2013}.

The main drawback of Eq. \eqref{eq:wavesym} is that, compared to a two-point average of the density or the kernel, the procedure of symmetrizing the wavevector is not physically transparent. Of course, two-point schemes also suffer from limitations (e.g.\ the kernel has no knowledge of the medium between $\mathbf{r}$ and $\mathbf{r'}$). The fact that we have to invoke any averaging system at all is an undesirable consequence of using kernels derived from the HEG to describe inhomogeneous systems. In what follows we present calculations using both the density and wavevector
symmetrization schemes.

\subsection{Computational details and convergence}
The kernels described in previous sections has been implemented in the DFT code GPAW\cite{Enkovaara2010a, Larsen2017}, which uses the projector augmented wave (PAW) method.\cite{blochl} The calculation of correlation energies energies in the framework of the ACFD are performed in four steps. 1) A standard LDA of PBE calculation is carried out in a plane wave basis. 2) The full plane wave Kohn-Sham Hamiltonian is diagonalized to obtain all unoccupied electronic states and eigenvalues. 3) A plane wave cutoff energy is chosen and the Kohn-Sham response function\cite{jun} is calculated by putting number of unoccupied bands equal to the number of plane waves defined by the cutoff. 4) The correlation energy is evaluated according to Eqs. \eqref{eq:E_c_x} and \eqref{eq:dyson}. The calculated correlation energies are finally added to non-selfconsistent Hartree-Fock energies evaluated on the same orbitals as the correlation energy. The coupling constant integration is evaluated using 8 Gauss-Legendre points and the frequency integration is performed with 16 Gauss-Legendre points with the highest point situated at 800 $eV$. 

\begin{figure}[t]
    \centering
	\includegraphics[width=8.5 cm]{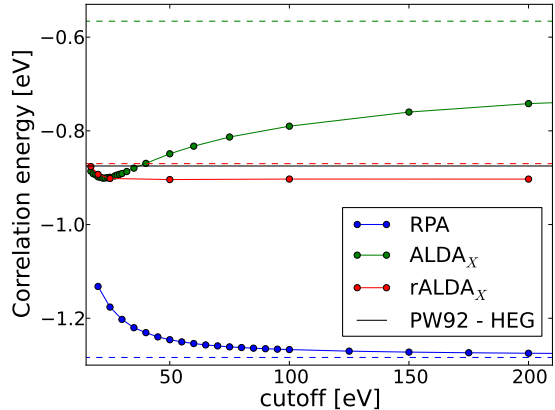} 
\caption{Correlation energy of the valence electron in Na evaluated with RPA, ALDA, and rALDA. The dashed lines show the values obtained with the functionals for the homogeneous electron gas using the average valence density of Na.}
\label{fig:Na}
\end{figure}

The main convergence parameter for these calculations is thus plane wave cutoff energy ($E_{cut}$) used for the response function and kernel. In the case of RPA calculations it has been shown that for sufficiently high cutoff energies the correlation energy scales as\cite{Harl2008,Olsen2013}
\begin{equation}\label{eq:extra}
 E_c(E_{cut})=E^c+\frac{A}{E_{cut}^{3/2}}
\end{equation}
and it is thus possible to perform accurate extrapolation to the converged results from a few calculations at low cutoff energies. When the ACFD method is used with a kernel the extrapolation \eqref{eq:extra} is less accurate, but the calculations often converge much faster than RPA such that extrapolation is either not needed at all or only introduce small errors. As an example we show the correlation energy of bulk Na in Fig. \ref{fig:Na} calculated with RPA, ALDA and rALDA. It is expected that the correlation energy should resemble that of a HEG with the average valence density of Na due to the delocalized valence electrons in Na \cite{miyake}. The rALDA calculations are rapidly converged with respect to unit cell sampling (two nearest unit cells are sufficient) and the result are shown in Fig. \ref{fig:Na} as a function of plane wave cutoff energy along with the RPA and ALDA results. Similarly to the HEG we find that RPA significantly underestimates the correlation energy while ALDA$_X$ overestimates it. We also note the very slow convergence of the ALDA calculation with respect to plane wave cutoff due to the $q$-independent kernel. 

\begin{figure}[tb]
    \centering
	\includegraphics[width=7.0 cm]{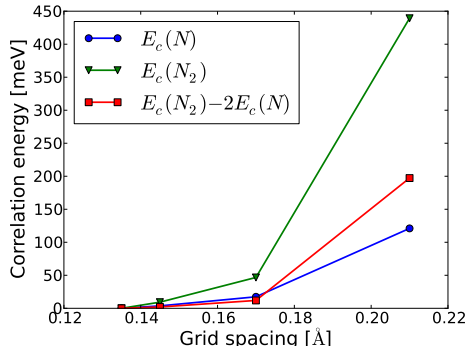}  
\caption{Convergence of rALDA correlation energy with respect to grid spacing for the atomization energy of N$_2$.}
\label{fig:N2_grid}
\end{figure}

In a plane wave representation the non-local kernels considered in the present work takes the form of Eq. \eqref{f_GG}. While the response function is calculated within the full PAW framework, it is not trivial to obtain the PAW corrections for a non-local functional. However, since the ALDA kernel vanishes for large densities, the non-local kernels considered in the present work tends to be small in the vicinity of the nuclei where it is usually difficult to represent the density accurately. As a consequence the kernels are rather smooth - even at the points where the density is non-analytical - and the kernel can be evaluated from the all-electron density represented on a uniform real-space grid using Eqs. \eqref{f_GG} and \eqref{f_tilde}. This is illustrated in Fig. \ref{fig:N2_grid} where the correlation energy of an N$_2$ molecule is shown as a function of grid spacing. The energy difference (contribution to the atomization energy) converges rapidly and is accurate to within 10 meV at 0.17 {\AA}. For the calculations in the present work the grid spacing was determined by the plane wave cutoff as $h=\pi/\sqrt{4E_{cut}}$ and a plane wave cutoff of 600 eV for the initial DFT calculations typically produces a grid spacing, which is $h\sim0.16$ {\AA}.

\section{Results}\label{sec:results}
\subsection{Absolute correlation energies}\label{sec.abscorr}
We have already shown that the RPA underestimates the correlation energy of the homogeneous electron gas by 0.6-0.3 eV per electron, whereas the ALDA overestimates the correlation energy by 0.3 eV per electron compared to the RPA. This is significantly improved by the rALDA functional, which gives an error of less than 0.05 eV (See Fig. \ref{fig:e_heg}). In Tab. \ref{tab:correlation}, we show that this trend remains true for simple atoms and molecules. For the H atom RPA gives a correlation energy of -13 kcal/mol (-0.56 eV) whereas ALDA gives 6 kcal/mol (0.26 eV). rALDA on the other hand gives -2 kcal/mol, which is a factor of three better than ALDA and a factor 6 better than RPA. A similiar picture emerges from the correlation energy of H$_2$ and the He atom.
\begin{table}[tb]
\begin{center}
\begin{tabular}{c|c|c|c|c|c|c}
\hline\hline      & LDA & PBE &  RPA & ALDA$_X$ & rALDA & Exact\\
	\hline
H     & -14 & -4  & -13 &   6  & -2   &   0 \\
H$_2$ & -59 & -27 & -51 & -16  & -28  & -26 \\
He    & -70 & -26 & -41 & -19  & -27  & -26 \\
\hline\hline
\end{tabular}
\end{center}
\caption{Correlation energies of H, H$_2$ and He evaluated with different functionals. Exact values are taken from Ref.~\cite{lee}. All numbers are in kcal/mol.}
\label{tab:correlation}
\end{table}

Although the RPA is free from self-interaction (cancellation of Hartree and exchange for single electron systems), it still contains a large amount of self-correlation, which originates from the fact that the Hartree kernel is not balanced by an exchange kernel in the ACFDT formalism. The self-correlation can be cancelled exactly by including second order screened exchange (SOSEX)\cite{Gruneis2009,Paier2010} or an exact exchange kernel\cite{hesselmann1,hesselmann2}. However, these approaches are far more computationally demanding than the ACFDT formalism. In contrast the rALDA kernel has a similar computational cost as RPA and reduces the self-correlation error to less than 0.1 eV for a H atom. The remaining error in rALDA is largely due to the choice of the LDA functional as the starting point and choosing the rAPBE kernel instead reduces the error to less than 1 meV.

\begin{figure}[tb]
        \includegraphics[width=\columnwidth]{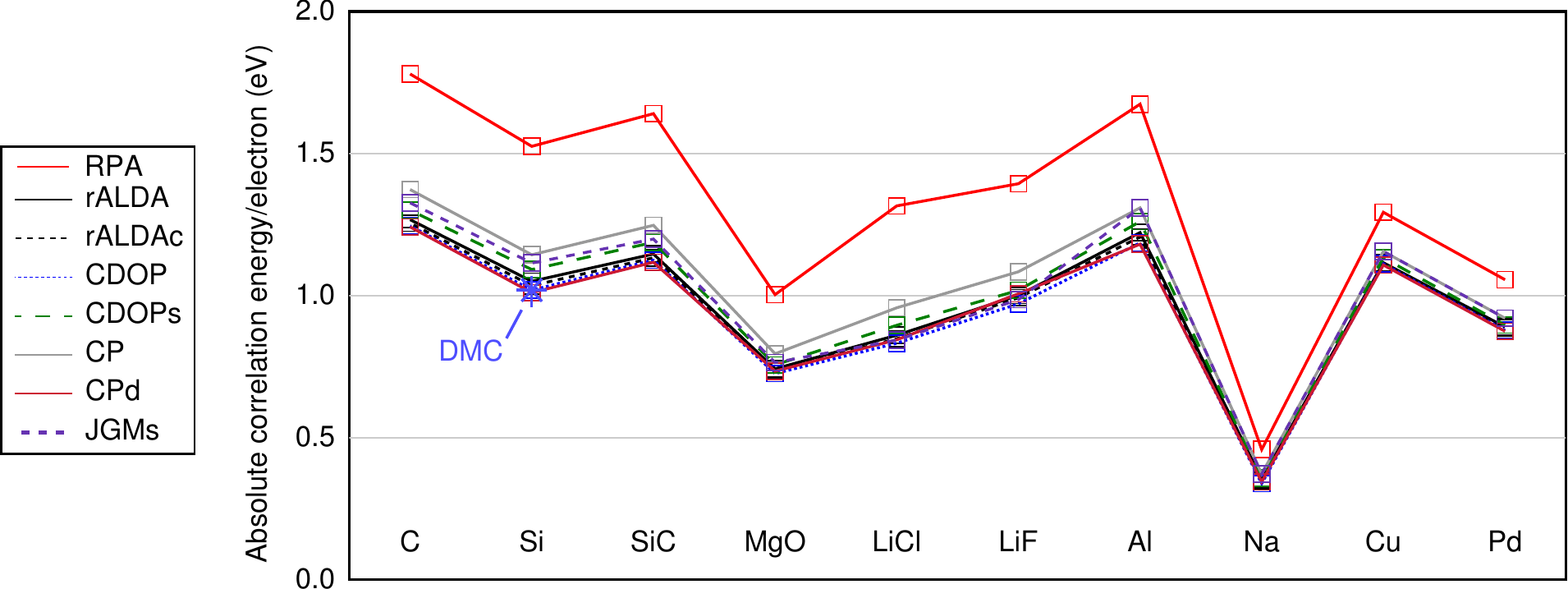}
\caption{Absolute correlation energy per valence electron calculated with different xc-kernels\cite{Patrick2015}.
Also shown is the correlation energy for Si obtained from diffusion Monte Carlo (DMC) calculations in Ref. \cite{Hood1998}.}
\label{fig:correnergy_materials}
\end{figure}

In Figure~\ref{fig:correnergy_materials} we show the correlation energy per valence electron
calculated for a number of solids using various  xc-kernels, and the RPA\cite{Patrick2015}.
As for the molecules and homogeneous electron gas, the RPA correlation energy is consistently larger (by
a few tenths of an eV/electron) than that calculated using the xc-kernels.
However, the differences in correlation energy between the kernels themselves are much smaller.
For instance, including the correlation contribution of the ALDA in the rALDA kernel (``rALDAc'')
increases the correlation energy by around just 1\% ($\sim$0.01 eV/electron) compared to the standard
rALDA.

Fig. \ref{fig:correnergy_materials} also shows the result of diffusion Monte Carlo (DMC) calculations
of the correlation energy of Si\cite{Hood1998}, which we can tentatively compare to our own results.
The DMC correlation energy lies among the values calculated from the kernels.
Ref. \cite{Lu2014} also found a correlation energy close to the DMC value using the CDOP
kernel and a different averaging scheme.
This result is of course reassuring, but we note that care must always be exercised when
making such comparisons.
First, one must expect the calculated value to have some dependence on the treatment of the
core-valence interaction (e.g.\ PAW, pseudopotentials, all-electron).
More generally, the concept of the correlation energy ``per valence electron'' becomes less
well-defined if the calculated correlation energy includes the contribution of semicore states.
For instance, comparing Figs.~\ref{fig:correnergy_materials} and~\ref{fig:Na} reveals an apparent
contradiction, that the valence energy per electron of Na in Fig.~\ref{fig:correnergy_materials}
is apparently approximately half its value in Fig.~\ref{fig:Na}.
This discrepancy is resolved, however, by noting that in the calculations of Fig.~\ref{fig:correnergy_materials}\cite{Patrick2015},
the entire 2$s^2$2$p^6$ shell of Na was included as valence in addition to the 3$s$ electron, unlike in Fig.~\ref{fig:Na}.
Therefore the correlation energy of Na reported in Fig.~\ref{fig:correnergy_materials} represents an
average over the free-electron-like $3s$ electron and the more localized $2s^22p^6$ shell,
which cannot be straightforwardly compared to the free electron gas as in Fig.~\ref{fig:Na}.

It is remarkable that the amount of self-correlation introduced by RPA is similar for widely different systems and it indicates that there will be a large energy cancellation when considering energy differences. As we will see below this is true to some extent, but the cancellation is far from perfect and RPA gives rise to systematic errors in cohesive energies of solids and atomization energies of molecules.

\subsubsection{Effect of kernel averaging scheme}
As discussed in Sec.~\ref{sec:symmetrization}, there is a choice in how one constructs the XC kernel for an inhomogeneous system. For instance, the correlation energies displayed in Table~\ref{tab:correlation} were calculated using a two-point average of the density. If we repeat the rALDA calculations using the symmetrized wavevector scheme (equation~\ref{eq:wavesym}) we obtain correlation energies of 6, -24 and -23 kcal/mol for H, H$_2$ and He, i.e.\ a difference of +8, +4 and +4 kcal/mol compared to the values of Table~\ref{tab:correlation}. The small differences in H and H$_2$ cancel when calculating the atomization energy\cite{Patrick2015}. We find it encouraging that the symmetrized wavevector approach agrees very well with the more intuitive two-point density average when calculating the atomization energy.

\subsection{Structural parameters}\label{sec:structure}
Having established that the renormalized kernels yield greatly improved absolute correlation energies, we now consider physical observables, starting with lattice constants and bulk moduli of a test set of 10 crystalline solids. In these calculations, the Kohn-Sham states and energies obtained self-consistently
within the LDA were used to calculate the noninteracting response function and exact exchange contribution to the total energy. A number of different XC kernels (including the rALDA) were used to calculate the response function and correlation energy through equations~\ref{eq:E_c_x} and \ref{eq:dyson}. The wavevector symmetrization scheme (equation~\ref{eq:wavesym}) was used to construct the kernel. The lattice constants and bulk moduli were extracted by calculating the total energy as a function of lattice spacing and fitting the results to a Birch-Murnaghan equation of state. More computational details are given in Ref. \cite{Patrick2015}.

\begin{figure*}[tb]
    \centering
	\includegraphics[width=14 cm]{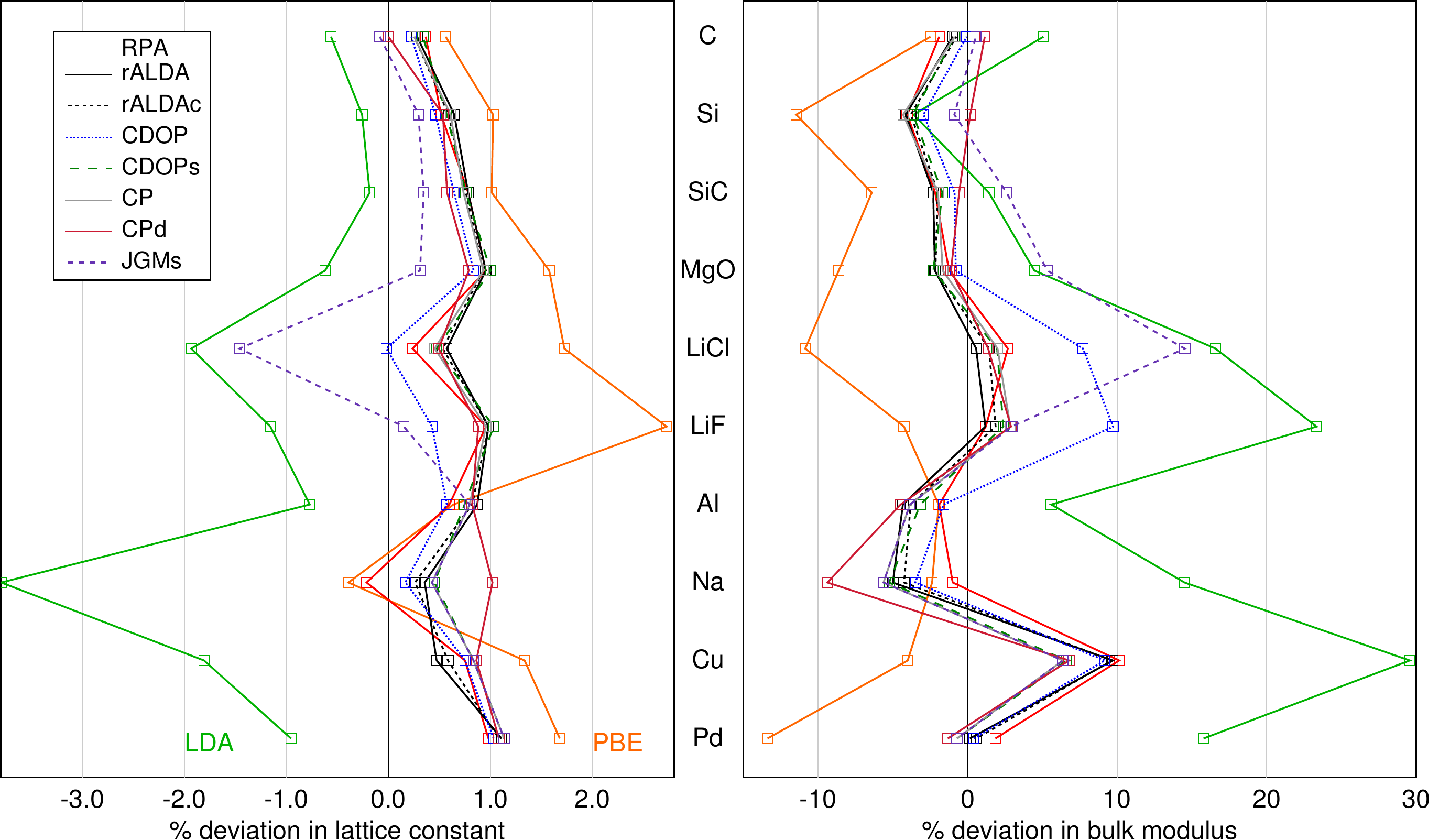} 
\caption{Lattice constants (left) and bulk moduli (right) calculated
using different approximations for the XC kernel
for a test set of 10 crystalline solids, compared to the experimental
values listed in Ref. \cite{Harl2010a}.  The experimental lattice
constants were corrected for zero-point motion\cite{Harl2010a}.
Note the CP and JGMs kernels coincide for metallic systems\cite{Patrick2015}.}
\label{fig:latt_BM}
\end{figure*}
Figure \ref{fig:latt_BM} shows the percentage deviation between the calculated structural parameters and the experimental data listed in Ref. \cite{Harl2010a}. Calculations performed within ground-state DFT within the LDA or GGA are also shown for comparison, which show the well-known tendency for the LDA/GGA to over/underbind, respectively. Using exact exchange and the ACFD correlation energy (with any kernel, or the RPA) systematically improves the agreement with experiment, going from a mean absolute error of 1.3\%/7\% for PBE to $\leq$0.7\%/4\% for lattice constants/bulk moduli, respectively.

The difference between the various kernels, and even the RPA, is rather small. In particular, the rALDA, rALDAc, CDOPs and CP kernels yield very similar results. The very close agreement between rALDA and rALDAc supports the use of the simpler rALDA kernel, which uses only the exchange part of the ALDA (Eq. \eqref{eq:rALDA}). One attractive property of the rALDA is that the calculated values display the fastest convergence with respect to the number of plane waves used to construct the response function, allowing a saving in computational time.

In terms of the other kernels, the strongest outlier is the JGMs (jellium-with-gap) kernel, particularly for the ionic solid LiCl. The agreement with experiment for the JGMs lattice constants can be improved even further by replacing the experimental optical gap $E_g$ that appears in the kernel definition with an effective gap inspired by excitonic calculations involving a long-range (LRC) attractive kernel\cite{Botti2004,Patrick2015}. One can also see that the CDOP kernel produces respectable structural parameters, despite it actually having a divergent pair-distribution function, while the variation between the CP and CPd kernels illustrate the potential importance of dynamical effects.

However, the overall differences between all of the kernels are rather small, and based on these calculations it is hard to argue that there are particular benefits in going beyond a simple, static kernel which tends to a density-dependent constant at small $q$ and decays as $1/q^2$ at large $q$. The rALDA satisfies these properties and also carries the particular advantage of scaling simply with the coupling constant $\lambda$. Finally, a significant strength of the rALDA is that, unlike the other kernels derived from the HEG, it has a spin-dependent generalization (Sec. \ref{sec:spin}), which is essential when calculating molecular atomization energies.

\subsection{Atomization energies of molecules}\label{sec.atoms}
\begin{figure}[tb]
\begin{center}
 \includegraphics[scale=0.4]{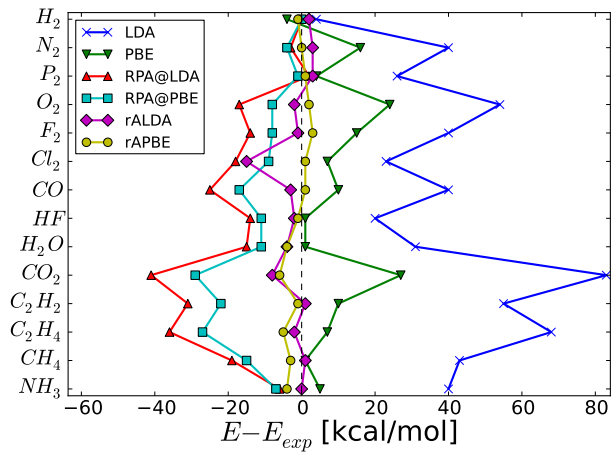}
 \caption{Molecular atomization energies evaluated with different methods shown relative to the experimental values. Results are shown with respect to reference values from Ref. \cite{karton}. The numbers are tabulated in the supplementary material of Ref. \cite{Olsen2014a}.}
\label{fig:energies_mol}
\end{center}
\end{figure}

In Fig. \ref{fig:energies_mol} we compare the performance of LDA, PBE, RPA@LDA, RPA@PBE, rALDA and rAPBE for the atomization energies of 14 small molecules using the experimental atomic positions. All numbers are shown relative to experimental values corrected for non-adiabatic effects\cite{karton}. RPA is seen to systematically underestimate the binding energies with RPA@LDA being slightly worse than RPA@PBE. In contrast LDA and PBE overestimate binding energies and the performance of PBE is similar to RPA whereas LDA is much worse. Both rALDA and rAPBE provides a significant improvement over RPA. rAPBE performs slightly better than rALDA, which is most likely due to the poor description of the ground state within LDA compared to PBE.

\begin{figure}[tb]
\begin{center}
 \includegraphics[angle=90, scale=0.4]{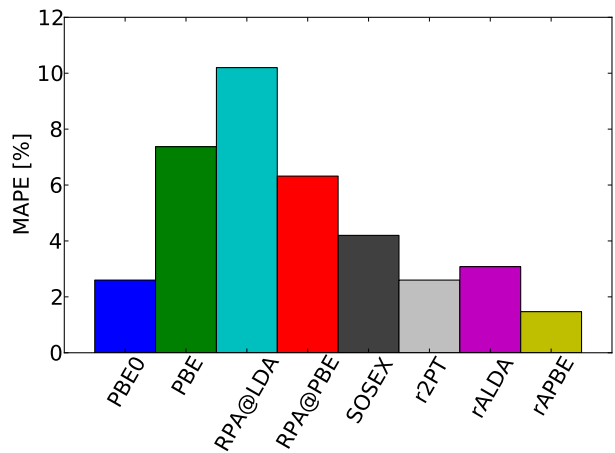}
 \caption{Mean absolute percentage deviation of molecular atomization energies. The PBE0, SOSEX, and rP2T values are taken from Ref. \cite{ren_review}.}
\label{fig:mape_mol}
\end{center}
\end{figure}

In Fig. \ref{fig:mape_mol} we show the mean absolute percentage error (MAPE) of RPA, rALDA and rAPBE compared with that obtained with PBE0 as well as SOSEX and rP2T, which constitute two other beyond RPA methods\cite{ren, Paier2012}. rALDA and rAPBE are a factor of three more accurate than RPA@LDA and RPA@PBE respectively. Moreover, the rAPBE MAPE is less than 1.5 {\%} and outperform both PBE0 and r2PT on this small test set.

\subsection{Cohesive energies of solids}\label{sec.solids}
While hybrid functionals may provide a computationally cheap way of obtaining accurate ground state energies for atoms and molecules, they typically fail dramatically for solids. Moreover, quantum chemistry methods are prohibitively demanding for solid state systems and DFT and DFT-based methods currently seem to be the only possible choice when dealing with solids. In Ref. \cite{Harl2010a}, it was shown that RPA performs somewhat worse than PBE for the cohesive energies of solids although it does provide significantly better results than LDA. This is in contrast to the case of molecules where RPA yields slightly better results than PBE. The reason is that the accuracy of RPA for atomization energies crucially depends on error cancellation of the ubiquitous self-correlation in RPA. The cancellation of errors is likely to work better when comparing similar systems, but for the cohesive energy of solids one has to consider ground state energies of atoms with ground state energies of solids, so the error-cancellation can become more inaccurate. Since the rALDA and rAPBE functionals to a large extent eliminate the self-correlation error of RPA, it is expected that these approaches should perform significantly better than RPA.

\begin{figure}[tb]
\begin{center}
 \includegraphics[scale=0.4]{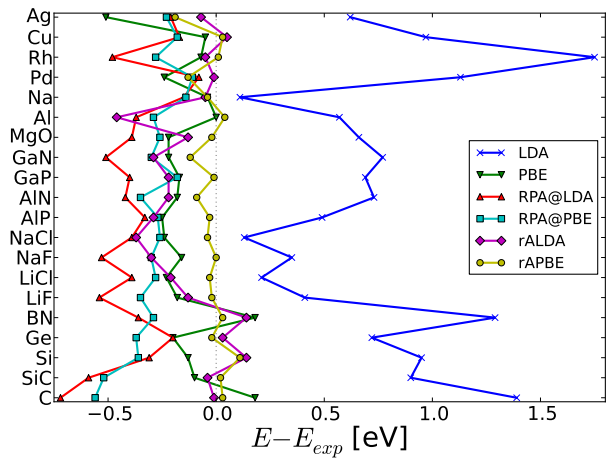}
 \caption{Deviation from experimental values of the cohesive energy of solids evaluated with different methods. The numbers are tabulated in the Supplemental Material of Ref. \cite{Olsen2014a}.}
\label{fig:energies_sol}
\end{center}
\end{figure}

\begin{figure}[tb]
\begin{center}
 \includegraphics[angle=90, scale=0.4]{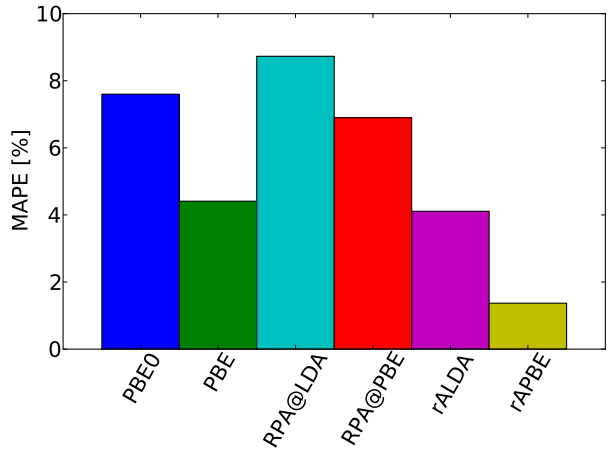}
 \caption{Mean absolute percentage deviation of cohesive energies of solids evaluated with six different methods.}
\label{fig:mape_sol}
\end{center}
\end{figure}

In Fig. \ref{fig:energies_sol} we show the cohesive energies of solids calculated with LDA, PBE, RPA, rALDA and rAPBE. Again the rAPBE functional performs significantly better than either PBE or RPA. The mean absolute percentage error is shown in Fig. \ref{fig:mape_sol} and the rAPBE deviation from experiment is less than 2 {\%}, whereas PBE, RPA@LDA and RPA@PBE give errors of 4 {\%}, 9 {\%} and 7 {\%} respectively. PBE0 gives a mean error of 7.5 {\%}, which is four times worse than rAPBE.

\subsection{Formation energies of metal oxides}
In the previous two sections we considered the problems of calculating the atomization energies of molecules and solids. This problem gauges the ability of a method to describe the absolute energy cost of breaking a chemical bond.
In most practical situations, however, it is often more relevant to consider the material's formation energy, i.e. its energy relative to the standard states of its constituent elements rather than the isolated atoms. The calculation of formation energies thus gauges the ability of a method to describe the energy of one type of chemical bond relative to another. 
Predicting the heat of formation of metal oxides has proven to be particularly challenging for a wide range of commonly applied xc-functionals. The RPA has previously been shown to significantly improve the accuracy of calculated formation energies of group I and II metal oxides as compared to semi-local functionals\cite{PhysRevB.87.075207}. In the following we briefly assess the performance of the rAPBE method and compare it PBE, RPA and the BEEF-vdW functional, which contains non-local correlation to account for van der Waals interactions\cite{PhysRevB.85.235149}.

The formation energy per oxygen atoms was obtained from the computed total energies as 
\begin{equation}
	\Delta E\mathrm{O} = \frac{1}{y}E[\mathrm{M}_x \mathrm{O}_y] - \frac{x}{y}E[\mathrm{M}] - \frac{1}{2}E[\mathrm{O}_2] \, , \label{eq:form_energy}
\end{equation}
where $E[\mathrm{M}_x \mathrm{O}_y]$, $E[\mathrm{M}]$ and $E[\mathrm{O}_2]$ are the total energies of the oxide, the bulk metal and the O$_2$ molecule in the gas phase respectively. Zero-point energy contributions were not included in the present study as previous work has shown that they affect the formation energies of oxides by less than 0.01 eV~\cite{PhysRevB.87.075207}.

\begin{table}
\centering
\begin{tabularx}{0.6\columnwidth}{lccccc}
\hline\hline \noalign{\smallskip}
 & PBE & BEEF-vdW & EXX & RPA & rAPBE  \\
\hline\noalign{\smallskip}
{\small MSE} & -0.55 & -0.40 & -0.96 & -0.38 & -0.18 \\
{\small MAE} & 0.55 & 0.40 & 0.99 & 0.38 & 0.21 \\
{\small MAPE} & 14.7\% & 10.9\% & 39.6\% & 12.1\% & 6.6\% \\
\hline\hline
\end{tabularx}
\caption{\label{tab:oxide_error} Mean signed error (MSE), mean absolute error (MAE), and mean absolute percentage error (MAPE) of calculated formation energies relative to experiments for 19 group I and II oxides and the transition metal oxides TiO$_2$ and RuO2$_2$. Energies are in eV per oxygen atom. Data taken from Ref. \cite{jauho2015improved}.}
\end{table}

\begin{figure}
    \centering
    \includegraphics[scale=0.5]{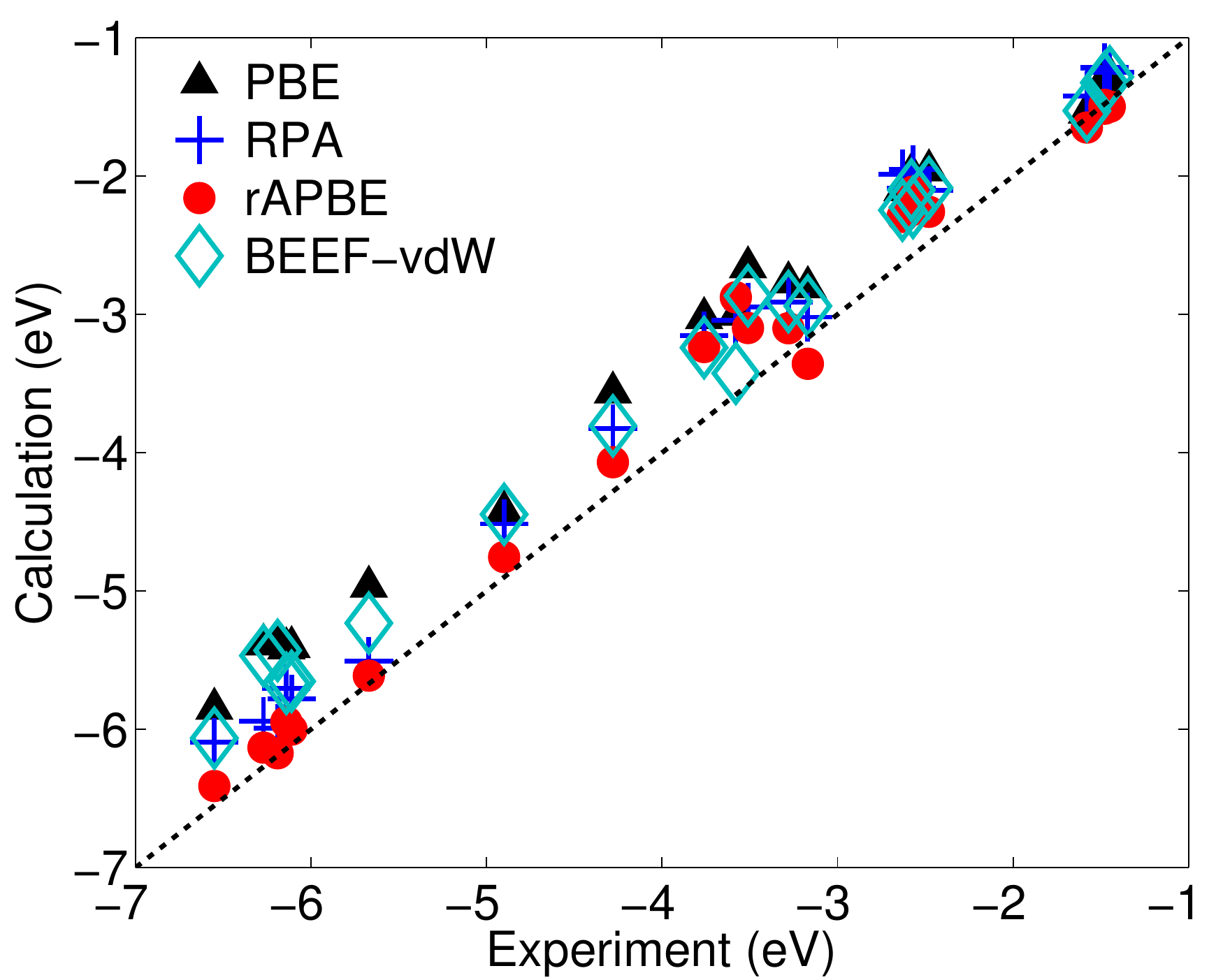}%
 \caption{Calculated oxide formation energy per oxygen atom using PBE, BEEF-vdW, RPA, and rAPBE plotted against the experimental data \label{fig:form_vs_exp}. The data set contains 19 group I and II metal oxides as well as the transition metal oxides TiO$_2$ and RuO$_2$. Data reproduced from Ref. \cite{jauho2015improved}.}
 \end{figure}

The formation energies computed with PBE, BEEF-vdW, EXX, RPA, and rAPBE are summarized in Fig. \ref{fig:form_vs_exp}. For the latter three methods, single-particle wave functions and energies were obtained from a self-consistent PBE calculation. All structures were optimized with PBE. The BEEF-vdW was included here to compare the performance of RPA and rAPBE methods to a semi-empirical method that explicitly includes dispersive interactions. The mean signed error (MSE), mean absolute error (MAE) and mean absolute percentage error (MAPE) with respect to experiment are shown Table \ref{tab:oxide_error}. Comparing the MSE and MAE shows that formation energies from PBE, BEEF-vdW, EXX and RPA have a clear systematic tendency to overstimate formation energies; that is  oxides are predicted less stable than found in experiments (see Ref.~\cite{PhysRevB.87.075207} for references for the experimental data). In contrast, while rAPBE shows the same tendency of destabilizing the metal oxide, it is less pronounced, and CaO$_2$, KO$_2$, CsO$_2$ and RuO$_2$ are in fact predicted to be more stable than experiment. The rAPBE method is in better agreement with experiments than all the other methods with a MAE of only 0.21 eV compared to 0.38 eV for RPA and 0.40 for BEEF-vdW. We can partly attribute the failure of the RPA to the lack of error cancellation between the correlation energy of the oxide and the bulk metal and oxygen molecule, which are all separately underestimated by the RPA (see Ref. \cite{jauho2015improved}). The errors of the DFT xc-functionals and the RPA are to some extent systematic and can be ascribed to a bad description of the O$_2$ molecule. In fact, treating the energy of the O$_2$ reference as a fitting parameter, the MAE for all the methods become comparable and lie in the range 0.15-0.2 eV/O\cite{jauho2015improved}. 

\subsection{Surface- and adsorption energies}
For applications of DFT to problems in surface science, in particular heterogeneous catalysis and electrocatalysis, the ability to predict stability and reactivity of metal surfaces is of crucial importance. It has been established that the RPA yields very good results for surface energies and chemisorption energies of atoms and small molecules on transition metal metal surfaces and greatly improves the accuracy of the xc-functionals \cite{schimka2010accurate,ren2009exploring,rohlfing2008binding,ma2011adsorption,kim2012benzene}. As shown below, the good performance of RPA for surface- and adsorption energies is preserved and probably even improved by the renormalized kernels. The difference between RPA and rALDA for surface reaction energies is on the order 5-10\%, which is comparable to the difference found for atomization energies of solids and molecules. This suggests that the better description of short range correlations by the kernel, which was found to improve atomization energies in Secs. \ref{sec.atoms} and \ref{sec.solids}, carries over to metal-molecule bonding. However, due to the significantly smaller magnitude of such bond energies compared to covalent bonds in solids/molecules and the lack of experimental data with sub 100 meV accuracy, it is not possible to confirm this hypothesis directly.

\begin{figure}[tb]
    \centering
	\includegraphics[width=8.0 cm]{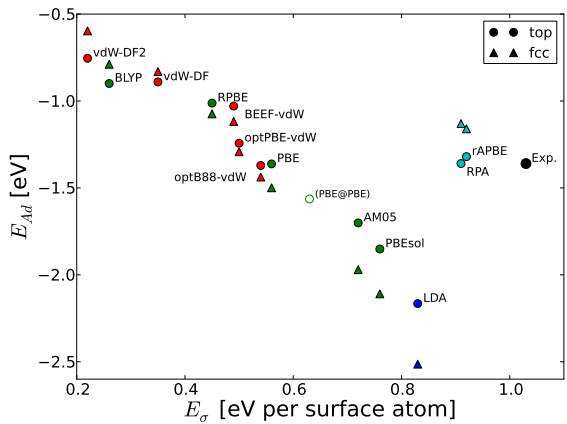} 
\caption{Surface energy versus adsorption energy of CO/Pt(111) calculated with various GGA functionals (green
markers) and van der Waals functionals (red markers). Circles and triangles indicate atop and hollow sites, respectively. All calculations were performed with the experimental lattice constant of Pt and the CO molecule relaxed with PBE. The hollow circle was obtained with a PBE optimized lattice constant. The coverage of CO is 1/4 and the Pt surface was modelled by a slab containing four atomic layers.}
\label{fig:COPt111}
\end{figure}

Figure \ref{fig:COPt111} shows the adsorption energy of CO on Pt(111) for a coverage of 1/4 plotted against the surface energy. Results obtained with RPA, rAPBE, and a range of xc-functionals are shown together with the experimental result states in Ref. \cite{schimka2010accurate}. It is evident that the RPA and rAPBE methods are able to break the incorrect correlation between adsorption energy and surface energy exhibited by the xc-functionals\cite{schimka2010accurate} and thereby yield an accurate description of both quantities simultaneously. Similar conclusions have recently been demonstrated for other adsorbates and surfaces\cite{schmidt2018benchmark}.

Table \ref{tab:RPA_rALDA} reports reaction energies in the full coverage limit for the set of benchmark surface reactions introduced in Ref. \cite{wellendorff2015benchmark} over the early 3$d$ transition metal surfaces. The absolute and relative deviation from the rALDA result is shown in the last two rows. It is interesting to note that the xc-functional yielding the best agreement with the rALDA (and RPA) is the RPBE, which is a revised version of the PBE fitted to experimental data on surface reactions like the ones considered here\cite{hammer1999improved}. 
The rALDA and RPA results are in good agreement, with the largest deviation occurring for OH adsorption (difference of 0.11 eV). This agrees well with results of Fig. \ref{fig:COPt111} for CO adsorption on Pt(111) and for graphene adsorbed on Ni(111) (see Fig. \ref{fig:graph_Ni}).  Although the absolute deviation between rALDA and RPA for the surface reactions is small (MAE of 40 meV), the relative deviation is in fact similar to that obtained for the atomization energies of molecules and solids. However, it is interesting to note that RPA shows a small but systematic tendency to overbind the adsorbates relative to rALDA. This is opposite to the trend observed for the atomization energies of molecules and solids where RPA underestimates the bond strength by around 0.3-0.5 eV/atom relative to rALDA (see Figs. \ref{fig:energies_mol} and \ref{fig:energies_sol}). For metals, the cohesive energy is also underestimated by RPA, but by somewhat smaller degree (MSE of -0.15 eV relative to rALDA). The average deviation between RPA and rALDA for the bond energy per atom is shown in Table \ref{tab:MAEMSE} for the set of materials in Figs. \ref{fig:energies_mol} and \ref{fig:energies_sol} grouped according to material type.

The fact that molecule-metal bonding shows an opposite trend compared to atomization energies can be explained as follows: for reactions involving the breaking of covalent bonds in the initial adsorbate molecule, e.g., for reactions of the type 1/2A2 - A/metal (reactions 1, 2, 6-8 in Table \ref{tab:RPA_rALDA}), RPA will overestimate the reaction energy because the A-A bond strength is underestimated more than the A-metal bond (the latter being weaker than the former). For pure adsorption reactions of the form A2 - A2/metal (reactions R3-R5), RPA will overestimate the adsorption energy because the reduction of the internal A-A bond upon adsorption is underestimated more than the A2/metal bond. In both cases the reason for the (slight) overestimation of the reaction energy can thus be traced to a larger underestimation by the RPA in describing pure covalent bonds compared to bonds with partial metallic character.

\begin{table*}
\centering
\begin{tabularx}{1.9\columnwidth}{ll|ccccccccc}
\hline\hline \noalign{\smallskip}
 Ads. & Surf. & rALDA & RPA & \hspace{0.2cm}LDA\hspace{0.2cm} & PBE & \hspace{0.2cm}RPBE & vdW-DF2 & BEEF-vdW & mBEEF & mBEEF-vdW\\
\hline\noalign{\smallskip}
H & Mn & 0.64 & 0.64 & 0.09 & 0.41 & 0.55 & 0.58 & 0.56 & 0.44 & 0.30 \\
O & Mn & -0.76 & -0.81 & -1.84 & -1.05 & -0.70 & -1.04 & -0.87 & -1.00 & -1.16 \\
N & Mn & 2.30 & 2.30 & 1.06 & 1.82 & 2.15 & 2.08 & 2.03 & 1.93 & 1.74 \\
N$_2$ & Mn & 0.75 & 0.74 & -0.63 & 0.50 & 1.04 & 0.83 & 0.70 & 0.55 & 0.10 \\
CO & Mn & -0.14 & -0.21 & -1.61 & -0.50 & 0.03 & 0.06 & -0.22 & -0.50 & -0.95 \\
NO & Mn & -0.96 & -0.99 & -2.87 & -1.50 & -0.96 & -1.06 & -1.19 & -1.33 & -1.68 \\
CH & Mn & 3.62 & 3.59 & 3.22 & 3.56 & 3.72 & 3.56 & 3.49 & 3.71 & 3.51 \\
OH & Mn & 1.36 & 1.24 & 0.46 & 1.12 & 1.44 & 0.98 & 1.15 & 1.14 & 0.95 \\
CO & Sc & -0.60 & -0.61 & -1.48 & -0.97 & -0.71 & -0.94 & -0.89 & -0.94 & -1.10 \\
CO & Ti & -0.77 & -0.81 & -1.68 & -1.00 & -0.66 & -0.80 & -0.90 & -1.04 & -1.32 \\
CO & V & -0.79 & -0.85 & -1.87 & -1.01 & -0.58 & -0.68 & -0.84 & -1.03 & -1.39 \\
CO & Cr & -0.34 & -0.38 & -1.75 & -0.74 & -0.25 & -0.27 & -0.50 & -0.77 & -1.17 \\
\hline\noalign{\smallskip}
\multicolumn{3}{l}{MAE (eV) vs. rALDA} & 0.04 & 1.10 & 0.31 & 0.12 & 0.16 & 0.15 & 0.28 & 0.54 \\
\multicolumn{3}{l}{MAE (\%) vs. rALDA} & 8 & 218 & 58 & 24 & 29 & 22 & 56 & 115 \\
\hline\hline
\end{tabularx}
\caption{\label{tab:RPA_rALDA}Adsorption energies (in eV) for a few reactions at full coverage calculated with rALDA, RPA and seven different DFT xc functionals. Data taken from Ref. \cite{schmidt2018benchmark}.}
\end{table*}

\begin{table}
\centering
\begin{tabularx}{0.7\columnwidth}{lcc|cc}
\hline\hline\noalign{\smallskip}
& \multicolumn{2}{l}{$E^{\mathrm{RPA}} - E^{\mathrm{rALDA}}$} & \multicolumn{2}{l}{\hspace{0.2cm}$E^{\mathrm{RPA}} - E^{\mathrm{Exp.}}$} \\
\hline 
 & MAE & \hspace{0.5cm}MSE\hspace{0.5cm} &\hspace{0.1cm} MAE & \hspace{0.5cm}MSE\hspace{0.5cm} \\
\hline
{\small Molecules} & 0.48 & -0.44 & 0.52 & -0.52 \\
{\small Gapped solids} & 0.30 & -0.30 & 0.43 & -0.43 \\
{\small Metals} & 0.18 & -0.15 & 0.24 & -0.24 \\
\hline\hline
\end{tabularx}
\caption{\label{tab:MAEMSE} Difference in atomization and cohesive energies between RPA, rALDA and experiments for three different types of materials. Data taken from Ref. \cite{schmidt2018benchmark}.}
\end{table}

\subsection{Static correlation}
\begin{figure}[tb]
    \centering
	\includegraphics[width=8.0 cm]{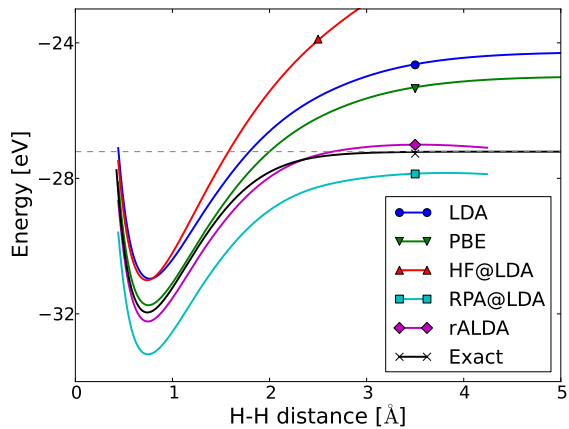} 
\caption{Dissociation curves of the H$_2$ molecule calculated with different functionals. The dashed line shows the energy of two isolated Hydrogen atoms (-1 Hartree). Each curve have been obtained by spline interpolation of 12 data points.}
\label{fig:H2}
\end{figure}

High highly attractive feature of the RPA is the correct description of bond dissociation in molecular dimers, which cannot be captured by ()restricted semi-local functionals. The bond dissociation curves in RPA are, however, only accurate of the reference energy of the isolated atoms are subtracted. In the case of H$_2$, for example, the dissociation curve is 1.1 eV below the exact result due to the self-correlation energy of the $H$ atom.
As we have seen, the rALDA large eliminates the self-correlation and one can obtain accurate dissociation curves of molecular dimers become without subtracting reference energies for the isolated atoms. This is shown in Fig. \ref{fig:H2}, where we also compare with LDA, PBE and Hartree-Fock, which all fail to yield the correct disociation energy. We note, however, that rALDA as well as rALDA exhibits a spurious maximum at intermediate bond lengths, which is not present in the exact result. The exact monotonous increase in energy signals a failure of both approximations. We note that is has previously been shown thatr the correct dissociation curve can be obtained from the ACDFT formalism if the interacting response function $\chi$ is evaluation from the Bethe-Salpater equation \cite{Olsen2014}. 

Finally it is worth mentioning that both RPA and rALDA fails dramatically for the dissociation of the H$_2^+$ dimer. In contrast, the SOSEX method yields exact result for this problem but cannot dissociate H$_2$ correctly.

\subsection{Dispersive interactions}
One of the main qualities of the RPA is its ability to account for van der Waals interactions in weakly interacting systems. Specifically, the RPA has been shown to provide an excellent description of the equilibrium geometry in hBN\cite{marini} and graphite\cite{lebegue} as well as graphite adsorbed on metal surfaces\cite{Olsen2011,mittendorfer,Olsen2013}. Conserving the accurate description of dispersive interactions is thus a major success criterion for any beyond-RPA method for ground state energies.

\subsubsection{Bilayer graphene}
\begin{figure}[tb]
\begin{center}
 \includegraphics[scale=0.4]{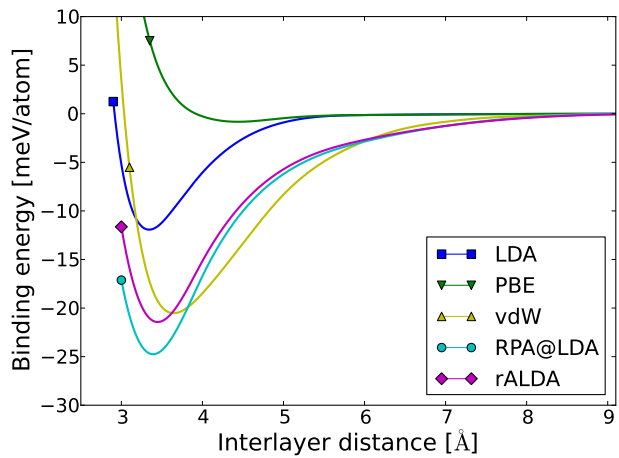}
 \caption{Binding energy of bilayer graphene calculated with RPA, rALDA as well as LDA, PBE and a van der Waals functional (vdW).}
\label{fig:bilayer}
\end{center}
\end{figure}

In Fig. \ref{fig:bilayer} we show the binding energy as a function of interlayer distance for a bilayer of graphene calculated with LDA, PBE, the van der Waals functional of Dion et al.\cite{dion} (vdW) along with the results of RPA and rALDA. The RPA and rALDA is seen to yield nearly identical equilibrium distances of 3.4 \AA, but rALDA gives a slightly smaller binding energy (22 meV per atom) compared to RPA (25 meV per atom). The semi-local functionals are not expected to capture dispersive interactions and PBE predicts a very shallow minimum at 4.4 \AA. However, LDA yields an equilibrium distance of 3.3 \AA, which is close to the RPA/rALDA value, but only half the binding energy. It is remarkable that LDA provides a seemingly qualitative correct description of the binding energy in this system. Similar results for LDA has previously been obtained for graphite\cite{lebegue} and graphene on metal surfaces\cite{Olsen2013}, but the results are likely to be fortuitous, and the fact that LDA and various GGAs produce qualitatively different results renders the result dubious.

The vdW functional predicts an equilibrium distance of 3.7 \AA and a binding energy of 22 meV per atom. In contrast to LDA, this functional binds the two layers for the right reasons, but the binding distance is quantitatively wrong. It is also noteworthy that the tails of the potential energy surfaces of RPA and rALDA coincide as expected, but deviate from the prediction of the vdW functional. It should be noted that there is no experimental values for the binding energy and binding distance of bilayer graphene, but it is expected that the binding distance should be close to the value of 3.34 {\AA} of graphite.

\subsubsection{Graphene on metals}
\begin{figure}[tb]
\centering
 \includegraphics[scale=0.5]{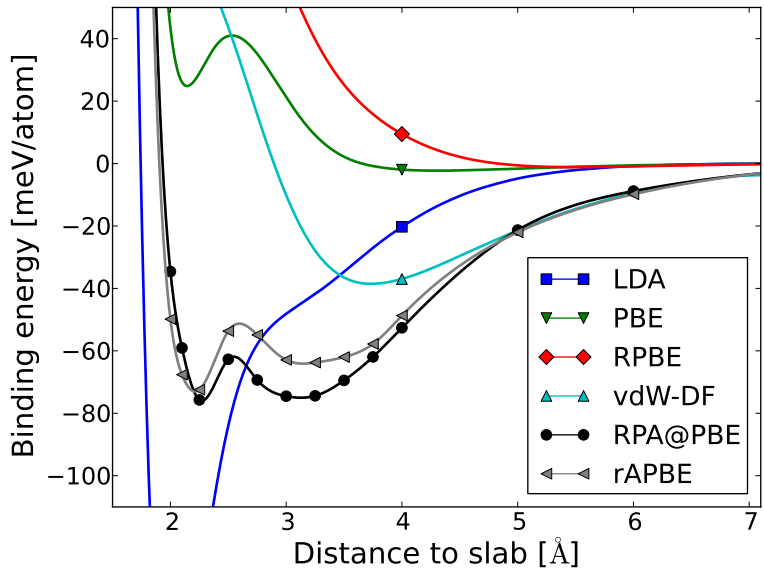}
 \caption{Binding energy of graphene on a Ni(111) surface calculated with RPA, rAPBE as well as LDA, PBE, RPBE and a vand der Waals functional (vdW-DF).}
\label{fig:graph_Ni}
\end{figure}

The description of graphene adsorbed on metal surfaces has proven a highly challenging task for first principles methods due to equal contributions of weak covalent and van der Waals bonding in these systems\cite{Olsen2011, mittendorfer}. In Fig. \ref{fig:graph_Ni} we compare the binding energy of graphene on a Ni(111) surface calculated with LDA, PBE, RPBE, vdW, RPA and rAPBE\cite{Olsen2014a}. The RPA shows two distinct minima at 2.2 {\AA} (chemisorbed) and 3.3 {\AA} (physisorbed), which originates from weak covalent interactions and dispersive forces respectively. The chemisorption minimum is a few meV lower than the physisorption minimum and agrees well with the experimentally determined value of 2.1 {\AA} for this system. The rAPBE functional closely mimics the RPA result, but gives a slightly larger energy difference between the two local minima. The vdW functional completely misses the weak covalent interactions responsible for the chemisorption minimum and only yields a physisorbed state at 3.8 {\AA}.

\subsubsection{C$_6$ coefficients}
Long-range interactions are typically dominated by Coulomb interactions and accuracy of dispersive interactions are thus determined by the response functions of the individual systems. For well-separated atoms the binding energy asymptotically becomes $E_B(r)=C_6/r^6$, where the $C_6$ coeffients are only depend on the polarizabilities of the isolated atoms. In particular the $C_6^{ij}$ coefficient relating atoms $i$ and $j$ can be calculated from the Casimir-Polder formula
\begin{equation}\label{eq:c6}
 C_6^{ij}=\frac{3}{\pi}\int_0^\infty\alpha_i(i\omega)\alpha_j(i\omega)d\omega,
\end{equation}
where  
\begin{equation}
 \alpha_i(i\omega)=-\int d\mathbf{r}d\mathbf{r'}z\chi_i(\mathbf{r},\mathbf{r'},i\omega)z'
\end{equation}
is the polarizability of atom $i$ and $\chi_i$ is the interacting response function, which can be calculated form the Dyson equation with a given approximation for the xc-kernel.

The $C_6$ coefficients for eight different atoms ($i=j$) is displayed in Tab. \ref{tab:c6} calculated with LDA ($\chi=\chi^{KS}$), RPA and rALDA. We observe that RPA performs significantly better than LDA, but rALDA is a factor of three better than RPA on average. The performance is, however, very dependent on the type of atom and rALDA performs better for the noble gas atoms, except for He, which is more accurately described in RPA. For Li and Na RPA fails completely, whereas LDA provides rather accurate predictions (better than rALDA for Li and slightly worse tahn rALDA for Na).
\begin{table}[tb]
\begin{center}
\begin{tabular}{c|c|c|c|c}   
\hline\hline     & LDA & RPA@LDA & rALDA & Exact \\
	\hline
He   &  2.2 & 1.5 & 1.8  & 1.44 \\
Ne   &    9 &   6 &   7  & 6.48 \\
Ar   &  140 &  57 &  67  & 63.6 \\
Kr   &  280 & 110 & 130  & 130  \\
Li   & 1290 & 493 & 1180 & 1380 \\
Na   & 1520 & 560 & 1280 & 1470 \\
Be   &  590 & 163 &  243 & 219  \\
Mg   & 1400 & 370 &  570 & 630  \\
\hline
MARE & 0.79 & 0.29 & 0.11  & \\
\hline\hline
\end{tabular}
\end{center}
\caption{C$_6$ coefficients between identical atoms ($i=j$ in Eq. \eqref{eq:c6}) calculated with LDA, RPA, and rALDA. All values are in atomic units. We also show the mean absolute relative error (MARE).}
\label{tab:c6}	
\end{table}

\subsection{Structural phase transitions}
Bulk solids usually exist in various polymorphic forms. Under changes in pressure or temperature, one structure may transition into another. The structural phase transition of solids has large theoretical and practical importance. With the external influence, the space-group symmetry and associated internal structural parameters change from one crystal structure to another. Temperature or pressure induced structural phase transitions can also change the electronic structures of the corresponding materials, such as from insulator to metal and vice versa, resulting in changes of the band-gap or conductance\cite{MRM03}. Structural phase transitions also sometimes lead to different magnetic states\cite{RIM97}. As a consequence, structural phase transitions offer an opportunity to tune a material toward particular applications in electronics, optics and other relevant fields\cite{VLM00,KNM03,KNL14}.

Structural phase transitions have largely remained a challenge for electronic structure methods. When computing phase transitions, a robust theory must overcome the dilemma of simultaneously predicting equilibrium structural properties and the phase transition parameters. Since experiments often fail to precisely measure the coexistence temperatures or pressures of two different structural phases of a solid, there is a high demand for a robust theoretical method. The failure of semilocal density functional approximations for structural phase transitions was earlier attributed to the underestimation of the band-gap with these methods\cite{BHH06,HWD10,MRM03,XSR13}. This assumption was later questioned by the results from the HSE (Heyd-Scuseria-Ernzerhof) approximation\cite{HSE03,HSE06}. HSE is nonlocal in the exchange and predicts more realistic fundamental gaps. HSE is better than semilocal functionals for the transition pressures of Si and SiO$_2$, but seriously overestimates the transition pressure in Zr.

In combination with the rAPBE and rALDA kernels, the RPAr approximation was investigated for the structural phase transitions of a small but representative group of materials\cite{SBR18}. The examples were chosen to incorporate several changes in band structure, including from semiconductor to semiconductor, semiconductor to metal, and metal to metal transitions. 
The assessment includes the phase transitions of the diamond phase of Si 
to the metallic beta-tin form, the zinc-blende (ZB) to rocksalt transition of SiC, the ZB to \textit{Cmcm} phase transition of GaAs, the quartz to stishovite transition of SiO$_2$, the transition from fcc to hcp structure of Pb, and finally the phase transition of the hexagonal to cubic structures of BN.

The transition pressure can be found as the negative slope of the common tangent line between the two phases. At the transition pressure, the difference in Gibbs free energies for the two phases should be equal to zero. 
At a finite, but constant, temperature,  the pressure can be calculated
as the negative derivative of the free energy with respect to the volume
\begin{align}
    P(V,T) = -\left(\frac{\partial F}{\partial V}\right)_T
\end{align}
At zero temperature the equivalent condition is that the enthalpy difference 
of the two phases is zero, and the pressure can be found directly as the derivative of the electronic energy.

To include the thermal effects, the equilibrium parameters were obtained from fitting the third-order Birch-Murnaghan equation of state (EOS) including the thermal corrections from the vibrational degrees of freedom. The results of the fitting are $F_{0T}$, the minimum of the Helmholtz energy at temperature $T$ and ($V_{0T}$, $B_{0T}$, $B'_{0T}$), the equilibrium parameters such as the equilibrium volume, bulk modulus and the derivative of the bulk modulus at that same temperature. The transition pressures were then obtained using the isothermal EOS fitting parameters obtained from Eq. (9) in Ref. \cite{SBR18}. Compared to zero Kelvin, the addition of thermal corrections introduces a rigid shift in the performance of all the methods for most of the materials in the assessment, and the agreement typically improves with experiment when the thermal corrections are included.

The results of different functionals for predicting the phase transitions generally follows the same trend as the structural parameters, Figure.~(1) in Ref.~\cite{SBR18}. In general, all beyond-RPA methods deliver an overestimation of the equilibrium and transition volumes by about the same amount as the local density approximation (LDA) underestimates them. Beyond-RPA approximations yielded an overall improvement compared to the bare RPA and semilocal functionals for the transition pressures, Table~\ref{tab:rpar-pt1}, although the behavior is less systematic than for RPA. 
\begin{table}
  \centering
  \begin{tabular}{lccccccc}
    Materials & LDA & PBE & SCAN &
    RPA & RPAr1 & HOT & Expt\cite{MRM03,SBR18} \\
    \hline
    Si & 6.3 & 8.5 & 13.8 & 12.8 & 10.4 & 9.7 & 12.0 \\
    Ge & 5.6 & 7.1 & 10.4 & 10.2 & 9.5 & 9.2 & 10.6 \\
    SiC & 56.4 & 61.4 & 69.1 & 69.6 & 66.8 & 65.6 & 100.0 \\
    GaAs & 9.4 & 11.6 & 16.1 & 18.0 & 16.3 & 16.0 & 15.0 \\
    SiO$_2$ & -0.1 & 6.4 & 5.2 & 4.3 & 7.1 & 7.4 & 7.5\cite{H96} \\
    Pb & 17.8 & 21.7 & 22.2 & 23.9 & 22.7 & 22.5 & 14.0 \\
    C  & 3.1 & 9.8 & 8.3 & 4.2 & 10.4 & 10.4 & 3.7 \\
    BN & 0.1 & 6.5 & 6.1 & 1.8 & 4.3 & 4.4 & 5.0 \\
  \end{tabular}
  \caption{Finite temperature transition pressures, in GPa, predicted at 300 K using different levels of density functional approximations\cite{SBR18}.
    The rAPBE kernel was used in combination with RPAr to obtain the bRPA results. RPAr1 is sufficient at capturing the needed bRPA correlation for predicting the phase transitions, and the HOT correctio introduces a small shift.}
\label{tab:rpar-pt1}
\end{table}

For Si and Ge the RPAr1 approximation delivers a reasonable result, however, RPA shows closer agreement with the experimental transition pressure. For SiC, neither RPA nor beyond-RPA show agreement with the experimental transition pressure, indicating that there is a significant difference between equilibrium and non-equilibrium transition pressures\cite{D07}. For SiO$_2$, and GaAs the transition pressure predicted with an xc kernel are more accurate compared to experiment than the bare RPA. Adding bRPA correlation from rAPBE at any level of RPAr reduces the transition pressure of RPA and comes quite close to the experiment. Ref. \cite{XSR12} 
attributes this deviation of RPA to its poor performance for some molecular-like solids where there is less cancellation of error between dissimilar phases. It is remarkable that RPA fails to predict the correct phase ordering for BN, while the rAPBE kernel in conjunction with RPAr 
brings the transition pressure close to the experimental value. The thermal corrections play a more prominent role in the phase transitions of carbon and BN, and indicate further investigations are necessary. RPA yields too low a transition pressure for the phase transition of carbon without thermal corrections, and reverses the phase stability of BN without a finite-temperature correction. All beyond-RPA approximations largely overestimate the transition pressure in C with the inclusion of thermal corrections, while the prediction for BN is significantly improved compard to RPA. The inclusion of the higher-order terms with the rAPBE kernel leads to a better accuracy against the experimental transition pressure in BN. The unexpected inaccuracies in materials such as Pb, C and BN, can be explained by near degeneracies.
\begin{figure}[tb]
    \centering
    \includegraphics[width=0.7\linewidth]{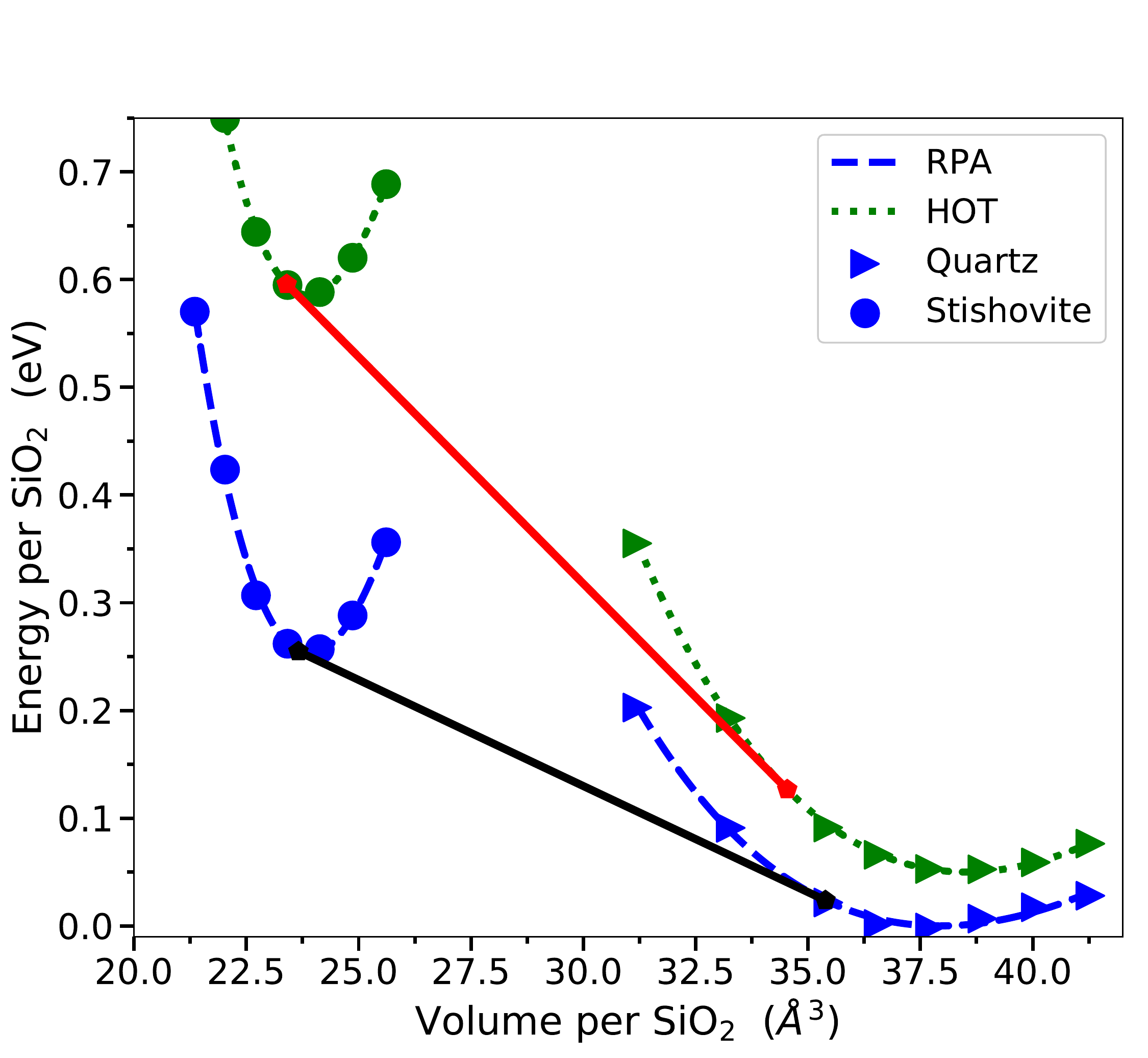}
    \caption{Energy-volume curves for the quartz and stishovite phases of SiO$_2$ with RPA and rAPBE (within the HOT approximation) per functional unit\cite{SBR18}. The negative slope of the common tangent line corresponds to the transition pressure. The kernel corrections for SiO$_2$ increase the equilibrium energy difference between phases and correct the large underestimation of the transition pressure by RPA 
    as a result. The kernel-corrected curves have been rigidly shifted up in energy by 0.05 eV compared to RPA for visual clarity.}
    \label{fig:sio2-pt}
\end{figure}

To illustrate the benefits of including the exchange-correlation kernel, 
Fig. \ref{fig:sio2-pt} shows the EOS data and common tangent results for SiO$_2$ predicted with RPA and the HOT approximation in combination with rAPBE. In this case, RPA underestimates the energy difference between the phases resulting in an underestimation of the transition pressure. The root cause of this can be understood if the transition pressure is thought of as $\Delta E/\Delta V$ evaluated at the transition volumes for each phase. The energy underestimation is more severe than the transition volume errors, and so RPA underestimates the pressure (as with most semilocal functionals). With the addition of the xc-kernel, the energy difference between phases is increased by an appropriate ammount to bring the rAPBE transition pressure within the experimental range.

\subsection{Cesium halide stability}
The difficulty in predicting the energy difference between similar phases of a material is a more general problem than phase transitions alone. The stability of different phases in alkali-halides is also a strong probe of various electronic structure methods and the correlation effects they incorporate. Among the alkali halides, those formed from cesium show an interesting behavior. CsF is experimentally stable in the B1 (rocksalt or NaCl) structure, while the Cl, Br, and I materials exist experimentally in the B2 (CsCl) structure. For all of the other alkali halides, the stable structure is B1. The stability of the B2 phase for certain cesium halides is a direct consequence of weak van der Waals bonding\cite{M33,L37,T63,R61,ZGU13,TZG17}. All ACFD-based methods 
naturally account for long-range van der Waals forces\cite{EF11}, but an accurate treatment of their structure requires both short and long-range interactions to be accounted for. Semilocal density functional approximations miss the long-range van der Waals forces. Ref \cite{ZGU13} indicated the relevance of long-range interactions through the PBE+D2 method compared to the bare PBE-GGA for the phase ordering of these cesium halides. Since D2 is a semi-empirical method\cite{G06}, the energy differences reported in Ref. \cite{ZGU13} are not necessarily usable as a benchmark. 

\begin{figure}[tb]
    \centering
    \includegraphics[width=0.8\linewidth]{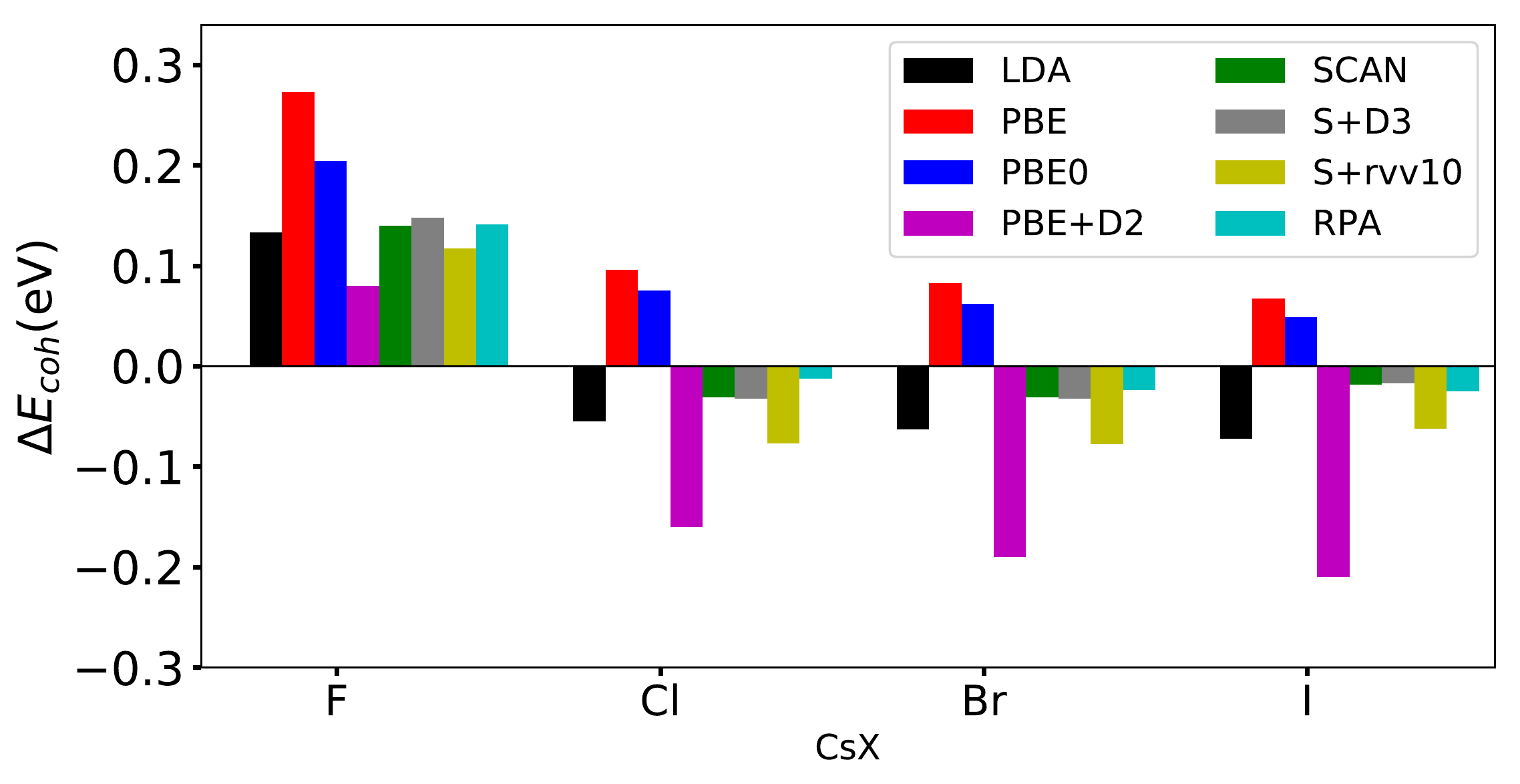}
    \caption{Bar graph summarizing the difference in cohesive energies, $\Delta E_{\text{coh}} = E^{\text{B1}}_\text{coh} - E_\text{coh}^{\text{B2}}$, obtained with various DFT methods. PBE+D2 results are taken from Ref. \cite{ZGU13}. Positive $\Delta E_{\text{coh}}$ corresponds to the B1 phase being preferred as the ground state,whereas negative values indicate the preferred stability of the B2 phase. PBE predicts all ground state cesium halides to be in the B1 phase whereas all other methods favor the B2 structure except in CsF. S+D3\cite{BBS16} and S+rVV10\cite{PYP16} correspond to the SCAN\cite{SCAN} semilocal results plus the dispersion method specified.}
    \label{fig:csx_stabil1}
\end{figure}

To go beyond the semilocal level, the energy differences between the B1 and B2 phases were explored using ACFD-based methods. The performance of the previously discussed ACFDT-based approximations was assessed in detail in 
Ref~\cite{NRB18} for the structural parameters and cohesive energies of cesium-halides. The rALDA kernel was used primarily for this study. Compared to semilocal density functionals, RPA yields superior structural parameters 
for all of the stable cesium-halides. Beyond-RPA approximations in combination with the rALDA kernel are in general even more accurate for predicting the lattice constants and bulk moduli. The predicted cohesive energies for all of these ACFDT methods are also more accurate compared to the PBE-GGA due to the incorporation of van der Waals interactions. RPA predicts the proper phase stability of all Cs halides whereas PBE only predicts the correct order for the fluoride, Fig. \ref{fig:csx_stabil1}.

By definition the cohesive energy includes the energy of the bulk and the 
constituent atoms. Describing accurately the energy of the bulk and free atoms simultaneously is a challenge for most electronic structure methods, and a method biased towards one paradigm will lead to inconsistent predictions of the cohesive energy. Depending on the kernel and level of RPAr approximation, the resulting methods yield different descriptions for the bulk and free atoms. Overall the cohesive energies of the stable Cs halides are more accurate with RPA for the F and Cl compounds than for the heavier halides. For Br and I, the rALDA kernel within RPAr improves the predicted cohesive energies compared to RPA. A possible explanation is that RPA is applied in a non-self-consistent manner with PBE input orbitals, which is not necessarily the best starting point for the more ionic X-F and Cl bonds. 
The addition of a kernel tends to improve the short-range correlation of the ACFDT and therefore compensates the error of the PBE orbitals that is more prominent in the cohesive energies for RPA for the larger Cs halides.
\begin{figure}[tb]\centering
    \includegraphics[width=0.8\linewidth]{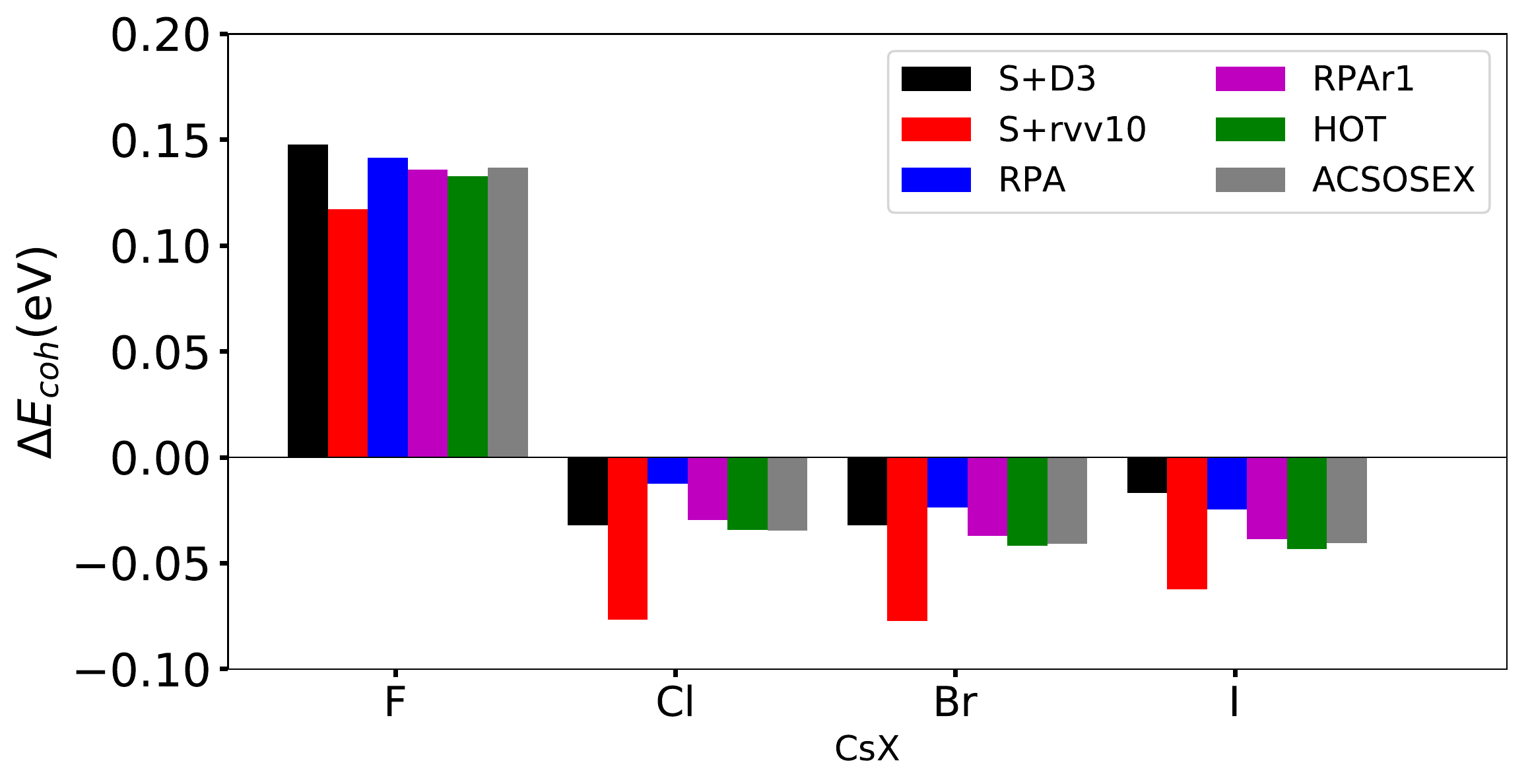}
    \caption{Bar graph representing the difference in cohesive energies, as in Fig.~\ref{fig:csx_stabil1}, obtained with beyond RPA methods using the rALDA kernel. The dispersion-corrected SCAN and RPA results are included for comparison. A positive $\Delta E_{\text{coh}}$ indicates the B1 phase is the preferred ground state, whereas negative values indicate the preferred stability of the B2 phase. SCAN plus dispersion gives a much more reasonable prediction compared to PBE+D2, if the ACFDT results are taken as the benchmark. }
    \label{fig:csx-stabil2}
\end{figure}

In order to correctly predict the splitting between the B1 and B2 phases a delicate balance of short and long-range correlation is required, Fig. \ref{fig:csx-stabil2}. The large overestimation of the stability difference between these two phases with the PBE+D2 method indicates the incompleteness of this approximation. RPA in contrast correctly incorporates long-range correlation, but is incomplete for the short-range part and so underestimates the energy difference. The energy difference is increased by any RPAr approximation using the rALDA kernel. rAPBE could be naturally expected to be the ideal kernel for evaluating the bRPA corrections on top of PBE orbitals, but interestingly this kernel struggles to predict the correct phase ordering in these materials. The other kernels tested, including rALDA, CP07\cite{CP07} and CDOP\cite{CDOP98}, all predict a consistent phase stability ordering, regardless of the RPAr approximation, further indicating the error lies in the rAPBE kernel itself\cite{NRB18}. This seems to be an isolated case, however, as the rest of the results in this review clearly demonstrate the utility and success of the rADFT kernels.

\subsection{Vertex corrected quasiparticle energies}\label{sec.vertex_results}
This section presents some results for quasiparticle energies obtained using the G$_0$W$_0$, GW$_0$, G$_0$W$_0$P$_0$, and G$_0$W$_0\Gamma_0$ self-energy methods outlined in Sec. \ref{sec:hedin}. For the latter two, the rALDA$_x$ xc-kernel was used and all calculations used an LDA starting point. The GW$_0$ refers to eigenvalue self-consistency in the Green function. All structures were relaxed using the PBE xc-functional and QP calculations were based on norm-conserving PAW potentials and spin-orbit coupling was included for the band structures. Further details on the calculations can be found in Ref. \cite{schmidt2017simple}. 

Table \ref{tab:res} reports the band gaps obtained with the different self-energy methods together with their mean absolute error (MAE) and mean signed error (MSE) relative to the experimental reference values for eight bulk crystals and three transition metal dichalcogenides in monolayer form. As expected G$_0$W$_0$@LDA underestimates the experimental band gaps. In agreement with previous findings, iterating to self-consistency in the Green's function (the GW$_0$ method) improves the situation somewhat, but leads to a small systematic overestimation of the gaps. This overestimation becomes even larger in fully self-consistent GW (not shown) where the MAE and MSE increase to 0.7 eV (see Ref. \cite{schmidt2017simple}). Including the rALDA vertex correction in a non-selfconsistent G$_0$W$_0\Gamma_0$ calculation, reduces the systematic underestimation of the gap somewhat (the MSE is reduced from -0.19 eV to -0.12 eV). Summarizing, we find that G$_0$W$_0\Gamma_0$ and GW$_0$ perform the best for the band gaps closely followed by G$_0$W$_0$. Including the vertex only in the polarizability (G$_0$W$_0$P$_0$) closes the gap because of the increased screening (as previously reported in Refs. \cite{shishkin2007m,chen2015accurate}) and results in a significant underestimation of the gaps.

\begin{figure*}[tb]
\centering
\includegraphics[width=0.9\linewidth]{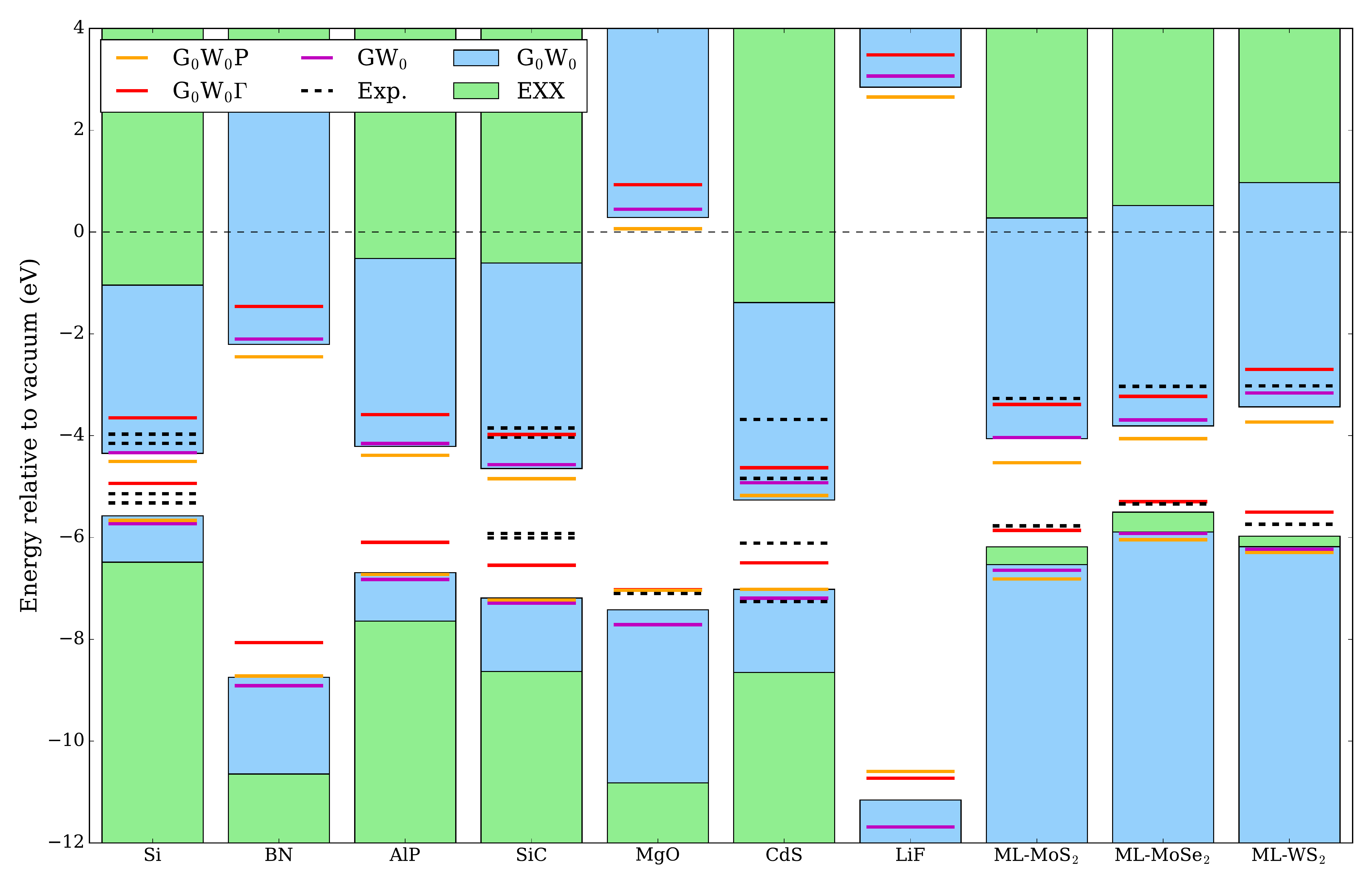}
    \caption{Absolute QP band positions relative to vacuum for a range of semiconductors and insulators as calculated with various methods. All calculations are performed non self-consistently from the LDA wave functions and eigenvalues. Experimental values are shown where available (black).}
    \label{fig:bands}
\end{figure*}

Fig. \ref{fig:bands} show the absolute band positions for the valence band maximum (VBM) and conduction band minimum (CBM) relative to vacuum. For the bulk materials the band positions were determined by aligning the Hartree potential of a bulk calculation with the potential in the center of a slab. The slab thickness was 10-24 atomic layers depending on the material and the surface orientation and reconstruction was taken from available experimental studies. The inclusion of the vertex has a striking effect of blue shifting the band edges by around 0.5 eV. Remarkably, this upshift yields a better overall agreement with experimental values (dashed black lines). Interestingly, this shift of band energies is not observed when the vertex is only included only in the polarizability. In addition, no systematic shift of the band edges is observed for the self-consistent GW approximations (without vertex corrections). We are thus led to conclude that the blue shift of band energies originates from the inclusion of vertex corrections in the self-energy.
\begin{table}[th]
\centering
\begin{tabular}{lcccccc}
\hline\hline \noalign{\smallskip}
& LDA & G$_0$W$_0$ & GW$_0$ & \multicolumn{1}{c}{G$_0$W$_0$P$_0$} & \multicolumn{1}{c}{G$_0$W$_0\Gamma_0$} & Exp.\\
\hline\noalign{\smallskip}
MgO & 4.68 & 7.70 & 8.16 & 7.10 & 7.96 & 7.98 \\
CdS & 0.86 & 1.76 & 2.27 & 1.84 & 1.87 & 2.48 \\
LiF & 8.83 & 14.00 & 14.75 & 13.25 & 14.21 & 14.66 \\
SiC & 1.31 & 2.54 & 2.72 & 2.38 & 2.57 & 2.51 \\
Si & 0.52 & 1.23 & 1.34 & 1.16 & 1.29 & 1.22 \\
C & 4.10 & 5.74 & 5.97 & 5.62 & 5.69 & 5.88 \\
BN & 4.36 & 6.54 & 6.81 & 6.27 & 6.60 & 6.60 \\
AlP & 1.44 & 2.48 & 2.67 & 2.34 & 2.51 & 2.47 \\
ML-MoS$_2$ & 1.71 & 2.47 & 2.61 & 2.28 & 2.47 & 2.50 \\
ML-MoSe$_2$ & 1.43 & 2.08 & 2.23 & 1.99 & 2.07 & 2.31 \\
ML-WS$_2$ & 1.33 & 2.75 & 3.07 & 2.56 & 2.81 & 2.72\\
\hline\noalign{\smallskip}
MAE & 1.89 & 0.20 & 0.17 & 0.41 & 0.16 & - \\
MSE & -1.89 & -0.19 & 0.12 & -0.41 & -0.12 & - \\
\hline\hline
\end{tabular}
\caption{Band gaps obtained using different self-energy approximations (see Sec. \ref{sec:hedin}) for eight bulk crystals and three monolayers. Experimental values for the bulk materials were corrected for zero-point motion and calculated band energies include spin-orbit coupling. The last two rows show the mean absolute error (MAE) and mean signed error (MSE) with respect to the experimental values. See Ref. \cite{schmidt2017simple} for more details.}\label{tab:res}
\end{table}

The physical origin of these effects can be traced to the improved description of short range correlations provided by the rALDA vertex. Indeed, the induced potential from Eq. \eqref{eq.RPA_W} is the Hartree potential generated by the induced density (the screening cloud of the quasiparticle). This Hartree potential is too deep and results in too deep-lying QP energies. This is the same reason underlying the systematic underestimation of the correlation energy by RPA. Adding the induced xc-potential by Eq. \eqref{eq.tildeW} reduces the size of the screening potential and thus shifts the QP energies up. As discussed in Ref. \cite{schmidt2017simple}, the band gap size is governed by long range correlations, which are well described by the RPA, while the \textit{absolute} band energies also depend on the short range correlations whose proper description require the vertex function. 

The observed upshift in QP energies by around 0.5 eV due to the vertex correction can also be understood by noting that the ionization potential and electron affinity, i.e. the VBM and CBM relative to vacuum, obtained from G$_0$W$_0$ (G$_0$W$_0\Gamma_0$) can be related to total energy differences between $N$ and $N\pm 1$ ground states evaluated from the ACFDT formula employing the RPA (rALDA) with a "frozen orbitals" assumption, i.e. a generalized Koopman's theorem\cite{niquet2004band}. Indeed, as we have shown in Sec. \ref{sec.abscorr} the RPA underestimates the absolute correlation energy by around 0.3-0.6 eV/electron and this error is largely repaired by rALDA due to the improved description of the short range correlations (see Fig. \ref{fig:kernels}). The incorrect description of absolute correlation energies by RPA largely cancels for energy differences when the states in question contain the same number of electrons. However, for QP energies where the initial and final states differ by $\pm 1$ electron, such errors are directly revealed.  

Finally, it should be mentioned that the suppression of the large $q$ components in the self-energy resulting from the rALDA kernel not only improves the description of local correlations but also leads to faster convergence with respect to the number of plane waves and unoccupied states, as compared to standard GW calculations\cite{schmidt2017simple}. The situation is very similar to that reported in Fig. \ref{fig:Na} for the ground state correlation energy in rALDA versus RPA and ALDA.

\section{Conclusions and outlook}\label{sec:conclusions}
We have reviewed the theory of exchange-correlation (xc-)kernels derived from the homogeneous electron gas (HEG), and illustrated how they can be used to obtain ground state correlation energies and quasiparticle band structures beyond the RPA and GW methods, respectively. While several xc-kernels have been introduced, we have mainly focused on the renormalized adiabatic kernels rALDA and rAPBE, which are obtained from the (semi)local LDA and PBE xc-functionals by a simple renormalization procedure. The renormalization procedure consists of a truncation of the large $q$-components of Fourier transformed xc-kernel, which renders it non-local in real space -- an essential property to avoid a divergent on-top correlation hole.   

The main observation is that the xc-kernels greatly improve the description of the short-range correlations relative to RPA (which correspond to setting $f_{xc}=0)$). In particular, the coupling constant averaged correlation hole, obtained from the density response function via the fluctuation dissipation theorem, becomes almost exact for the HEG and reduces the error on the correlation energy from about 0.5 eV/electron in the RPA to below 0.05 eV/electron. This improvement was shown to carry over from the extreme limit of delocalized electrons in the HEG to limit of localized states of simple atoms and molecules. For example, the spurious self correlation energy of the hydrogen atoms is reduced from 0.56 eV (RPA) to 0.04 eV (rALDA). The much better reproduction of absolute correlation energies reduces the reliance on error cancellation effects when computing energy differences. For RPA, error cancellation is very significant on the order of 0.5 eV/electron. For this reason RPA performs well for isoelectronic problems, problems where the electronic structure of the initial and final states are similar, e.g. for the calculation of structural parameters or the breaking/formation of weak (dispersive) bonds. When the xc-kernels are invoked the need for error cancellation is reduced, and as a consequence covalent bond energies, for which short-range correlations play an important role, are much more accurately described. This was explicitly demonstrated for atomization energies of molecules, cohesive energies of solids, formation energies of metal oxides, surface energies, and chemisorption energies of atoms and molecules on metal surfaces. In all these cases, the xc-kernels cure the systematic underbinding by the RPA and significantly improve the agreement with experimental data. Importantly, because the xc-kernels are of short-range nature they do not affect the shape of the correlation hole at longer distances, and consequently the excellent performance of the RPA for van der Waals interactions is preserved or even improved. For example the mean absolute error on the $C_6$ coefficients of noble gas and alkali elements are reduced from 0.3 eV (RPA) to 0.1 eV (rALDA). Similar conclusions apply to bonding with intermediate correlation ranges such as the mixed covalent-dispersive forces governing organic-metal interfaces as exemplified here by the prototypical graphene/metal interfaces. Furthermore, the positive effects that the xc-kernels bring can be adequately captured using RPA renormalized perturbation theory to first order, demonstrating that the impact of higher-order correlation effects are less important for correcting the shortcomings of RPA for structural and energetic properties.

In the context of quasiparticle band structure calculations within the formal framework of Hedin's equations, we have shown that the renormalized xc-kernels can be used to include vertex corrections in the electron self-energy. When the four-point kernel of many-body perturbation theory is approximated by the two-point kernel of time-dependent DFT, the effect of the vertex corrections attains a simple and physically transparent form. Namely, the effect of the vertex is to include the xc-potential into the dynamically screened Coulomb potential. We have shown that this is crucial in order to obtain a correct description of absolute band energies relative to the vacuum level. The effect of the xc-potential is to reduce the magnitude of the attractive QP screening cloud (in GW only the Hartree potential of the screening cloud is accounted for) and this lead to a significant up-shift of 0.3-0.5 eV for all energy levels leaving the band gap unchanged to within 0.2 eV. These effects should be important to include in order to correctly describe band alignment at surfaces and interfaces. 

While the adiabatic xc-kernels discussed in this review provide an improved description of the short range correlations leading to the many positive derived effects described above, they still present shortcomings. Most importantly, they do not provide any improvements for strongly correlated systems such as Mott insulators. For such systems it might be necessary to include frequency dependence or develop kernels with a more sophisticated density dependence than the HEG based kernels considered here. Another short-coming is the failure to reproduce the correct wavevector-dependence in the limit of long wavelengths. This is not an important issue for total energy calculations, but for optical absorption spectre it is crucial to have the right limiting behaviour in order to capture excitonic effects. The deficiency stems from the fact that the renormalization scheme is based on the correlation hole of the homogeneous electron gas and a proper treatment of the long wavelength limit in insulators is likely to require a scheme, which is not solely based on the density. One possible route towards this is to note that the truncation factor $\theta(2k_F-q)$ is the Fourier transform of the density matrix of a homogeneous electron gas with Fermi wavevector $2k_F$. In contrast to the bare density (defined by $k_F$), the density matrix provides a qualitative distinction between insulators and metals, but a direct truncation scheme based on the density matrix is not straightforward. Finally, the issue of self-consistency deserves a comment. 
In principle it is possible to define an optimized local effective potential (OEP) from the ACDFT with renormalized kernels, which may then form the basis of self-consistent solutions in the framework of the Kohn-Sham equations. However, the non-selfconsistent calculations are already rather computationally demanding and the OEP method does not seem to be a viable route to follow. In addition, the adiabatic kernels considered in the present work cannot be derived from any known ground state energy functional, which implies that it is somewhat inconsistent to evaluate the rALDA energy based on LDA orbitals for example. On the other hand, the situation is certainly improved compared to RPA where the Hartree orbitals comprises the only natural choice, but it would be highly desirable to have a class of xc-energy functionals that yield the renormalized kernels as their second functional derivative with respect to the density. A possible step in this direction is to define an LDA functional where the input density is replaced 
by a density convoluted with the Fourier transform of the truncation factor  $\theta(2k_F-q)$. This yield a ground an xc-energy functional, which leads directly to rALDA kernel if the density dependence of $k_F$ is neglected\cite{Olsen2014a}. Although such a scheme is approximate, the construction itself provides an intriguing new route to the development of novel xc-energy functionals that incorporate non-local effects through weighted density averaging.

\section{Acknowledgements}
KST acknowledges funding from the European Research Council (ERC) under the European Union's Horizon 2020 research and innovation program (grant agreement No 773122, "LIMA"). The work of AR was supported by National Science Foundation under Grant No. DMR-1553022. JEB acknowledges the A.R. Smith Department of Chemistry and Fermentation Sciences for support.
Figs.~\ref{fig:kernels}, \ref{fig:correnergy_materials} and  \ref{fig:latt_BM} are reprinted from J.\ Chem.\ Phys.\ 143, 102802 (2015), with the permission of AIP publishing.

\bibliographystyle{unsrt}


\end{document}